\documentclass[10pt]{article}

\usepackage[preprint]{tmlr}

\usepackage[utf8]{inputenc} 
\usepackage[T1]{fontenc}    
\usepackage{hyperref}  
\usepackage{url}            
\usepackage{booktabs}       
\usepackage{amsfonts}       
\usepackage{nicefrac}       
\usepackage{microtype}      
\usepackage{xspace}
\usepackage{graphicx}
\usepackage{subcaption}
\usepackage{makecell}
\usepackage{wrapfig}
\usepackage{soul}  
\usepackage{array}
\usepackage{ifthen}
\usepackage{siunitx}
\usepackage{balance}  
\usepackage{flushend}  
\usepackage{amssymb}
\usepackage{pifont}
\usepackage{multirow}
\usepackage{cleveref}

\newcolumntype{L}[1]{>{\raggedright\let\newline\\\arraybackslash\hspace{0pt}}m{#1}}
\newcolumntype{C}[1]{>{\centering\let\newline\\\arraybackslash\hspace{0pt}}m{#1}}
\newcolumntype{R}[1]{>{\raggedleft\let\newline\\\arraybackslash\hspace{0pt}}m{#1}}
\makeatletter 
\newcommand{\thickhline}{%
    \noalign {\ifnum 0=`}\fi \hrule height 1pt
    \futurelet \reserved@a \@xhline
}
\makeatother

\newcommand{\model}{\textsc{LLM~Compiler~FTD}\xspace}
\newcommand{\foundationmodel}{\textsc{LLM~Compiler}\xspace}
\newcommand{\codellama}{\textsc{Code~Llama}\xspace}
\newcommand{\codellamainst}{\textsc{Code~Llama~-~Instruct}\xspace}

\newcommand{\cmark}{\ding{51}}%
\newcommand\extrafootertext[1]{%
    \bgroup
    \renewcommand\thefootnote{\fnsymbol{footnote}}%
    \renewcommand\thempfootnote{\fnsymbol{mpfootnote}}%
    \footnotetext[0]{#1}%
    \egroup
}

\usepackage{listings}
\usepackage{xcolor}
\usepackage{float}
\newfloat{lstfloat}{htbp}{lop}
\floatname{lstfloat}{Listing}
\definecolor{commentsColor}{rgb}{0.497495, 0.497587, 0.497464}
\definecolor{keywordsColor}{rgb}{0.000000, 0.000000, 0.635294}
\definecolor{stringColor}{rgb}{0.558215, 0.000000, 0.135316}
\lstdefinestyle{mystyle}{
    basicstyle=\scriptsize\ttfamily,
    captionpos=b,
    breaklines=true,
    breakindent=0.5em,
    postbreak=\raisebox{0ex}[0ex][0ex]{\ensuremath{\color{red}\hookrightarrow}},  
    tabsize=2,
    frame=b,
    showstringspaces=false,
    numberstyle=\tiny\color{commentsColor},
    rulecolor=\color{black},
    commentstyle=\color{commentsColor}\textit,
    stringstyle=\color{stringColor},
    keywordstyle=\color{keywordsColor},
    emphstyle=\color{keywordsColor},
    escapeinside={(*@}{@*)},
}
\AtBeginDocument{\DeclareCaptionSubType{lstlisting}}

\title{Meta Large Language Model Compiler: Foundation Models of Compiler Optimization}
\author{\name%
Chris Cummins$^\dagger$,
Volker Seeker$^\dagger$,
Dejan Grubisic,
Baptiste Rozi\`{e}re,
Jonas Gehring,
Gabriel Synnaeve,
Hugh Leather$^\dagger$
\\ \\
\hspace*{0pt}\hfill Meta AI
}

\begin{document}

\maketitle
\extrafootertext{$\dagger$: Core contributors.}

\begin{abstract}
Large Language Models (LLMs) have demonstrated remarkable capabilities across a variety of software engineering and coding tasks. However, their application in the domain of code and compiler optimization remains underexplored. Training LLMs is resource-intensive, requiring substantial GPU hours and extensive data collection, which can be prohibitive. To address this gap, we introduce Meta Large Language Model Compiler (\foundationmodel), a suite of robust, openly available, pre-trained models specifically designed for code optimization tasks. Built on the foundation of \codellama, \foundationmodel enhances the understanding of compiler intermediate representations (IRs), assembly language, and optimization techniques. The model has been trained on a vast corpus of 546 billion tokens of LLVM-IR and assembly code and has undergone instruction fine-tuning to interpret compiler behavior. \foundationmodel is released under a bespoke commercial license to allow wide reuse and is available in two sizes: 7 billion and 13 billion parameters. We also present fine-tuned versions of the model, demonstrating its enhanced capabilities in optimizing code size and disassembling from x86\_64 and ARM assembly back into LLVM-IR. These achieve 77\% of the optimising potential of an autotuning search, and 45\% disassembly round trip (14\% exact match). This release aims to provide a scalable, cost-effective foundation for further research and development in compiler optimization by both academic researchers and industry practitioners.

\end{abstract}

\section{Introduction}

There is increasing interest in large language models (LLMs) for software engineering tasks including code generation, code translation, and code testing. Models such as StarCoder~\citep{starcoder2}, \codellama~\citep{codellama}, and GPT-4~\citep{openai2023gpt4} have a good statistical understanding of code and can suggest likely completions for unfinished code, making them useful for editing and creating software. However, there is little emphasis on training specifically to optimize code. Publicly available LLMs can be prompted to make minor tweaks to a program such as tagging variables to be stored as registers, and will even attempt more substantial optimizations like vectorization, though they easily become confused and make mistakes, frequently resulting in incorrect code.

Prior works on machine learning-guided code optimization have used a range of representations from hand-built features~\citep{wang2018machine} to graph neural networks (GNNs)~\citep{coreset}. However, in all cases, the way the input program is represented to the machine learning algorithm is incomplete, losing some information along the way. For example, \citet{mlgo} use numeric features to provide hints for function inlining, but cannot faithfully reproduce the call graph or control flow. \citet{programl} form graphs of the program to pass to a GNN, but exclude the values of constants and some type information which prevents reproducing instructions with fidelity. 

In contrast, LLMs can accept source programs, as is, with a complete, lossless representation. Using text as the input and output representation for a machine learning optimizer has desirable properties: text is a universal, portable, and accessible interface, and unlike prior approaches is not specialized to any particular task. 

However, training LLMs incurs high cost in both compute and data. For example, training \codellama's models consumed 1.4M A100 GPU hours to train, and curating the vast amounts of training data (hundreds of billions of tokens) can be challenging. These costs are often prohibitive to researchers in the field and this blocks advances that might otherwise be possible.

To address this issue, we are releasing \foundationmodel, a family of foundation models that have already been trained to understand the semantics of compiler IRs and assemblies and to emulate the compiler, allowing for easy fine-tuning with minimal data for specific downstream compiler optimization tasks. Building upon \codellama, we extend its capabilities to encompass compiler optimization and reasoning.

\begin{figure}
\begin{center}
\centerline{\includegraphics[width=\textwidth]{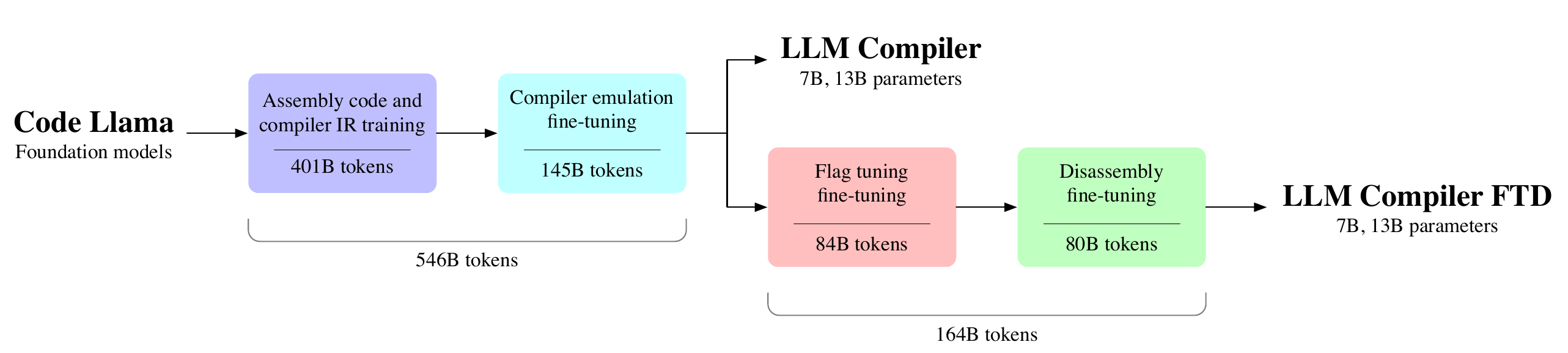}}
\caption{%
    \foundationmodel models are specialized from \codellama by training on 546 billion tokens of compiler-centric data in two stages. In the first stage the models are trained predominantly on unlabelled compiler IRs and assembly code. In the next stage the models are instruction fine-tuned to predict the output and effect of optimizations. \model models are then further fine-tuned on 164 billion tokens of downstream flag tuning and disassembly task datasets, for a total of 710 billion training tokens. During each of the four stages of training, 15\% of data from the previous tasks is retained.
}
\label{fig:training-pipeline}
\end{center}
\end{figure}

The training pipeline for \foundationmodel is illustrated in Figure~\ref{fig:training-pipeline}. We extend \codellama with additional pretraining on a vast corpus of assembly codes and compiler IRs, and then instruction fine-tune on a bespoke \emph{compiler emulation} dataset to better reason about code optimization. Our intention with releasing these models is to provide a foundation for researchers and industry practitioners to further develop code optimization models. We then adapt the models for two downstream compilation tasks: tuning compiler flags to optimize for code size, and disassembling x86\_64 and ARM assembly to LLVM-IR. We also release these \model models to the community under the same bespoke commercial license. Compared to the autotuning technique on which it was trained, \model achieves 77\% of the optimizing potential without the need for any additional compilations. When disassembling, \model creates correct disassembly 14\% of the time. On both tasks \model models significantly outperform comparable LLMs \codellama and GPT-4 Turbo.

Our work aims to establish a scalable, cost-effective foundation for further research and development in compiler optimization, catering to both academic researchers and industry practitioners. By providing access to pre-trained models in two sizes (7 billion and 13 billion parameters) and demonstrating their effectiveness through fine-tuned versions, LLM Compiler paves the way for exploring the untapped potential of LLMs in the realm of code and compiler optimization.

\subsection{Overview}

Figure~\ref{fig:training-pipeline} shows an overview of our approach. \foundationmodel models target compiler optimization. They are available in two model sizes: 7B and 13B parameters. The \foundationmodel models are initialized with \codellama model weights of the corresponding size and trained on an additional 546B tokens of data comprising mostly compiler intermediate representations and assembly code. We then further train \model models using an additional 164B tokens of data for two downstream compilation tasks: flag tuning and disassembly. At all stages of training a small amount of code and natural language data from previous stages is used to help retain the capabilities of the base \codellama model.
\section{\foundationmodel: Specializing \codellama for compiler optimization}
\label{sec:method}

\small
\begin{table}[t]
    \centering
    \setlength{\tabcolsep}{3pt}
    \begin{tabular}{@{}l@{}ccr@{}}
    \toprule
  Dataset &  Sampling prop. & Epochs &  Disk size \\  %
  \midrule
\multicolumn{4}{l}{\textbf{IR and assembly pretraining (401 billion tokens)}} \\ 
  Code     & 85.00\%  & 1.000 & 872~GB \\
  Natural language related to code             & 14.00\%  & 0.019 & 942~GB \\
  Natural language    & 1.00\%  & 0.001 & 938~GB \\
  \midrule
\multicolumn{4}{l}{\textbf{Compiler emulation (additional 145 billion tokens)}} \\ 
  Compiler emulation & 85.00\%  & 1.702 & 175~GB \\ 
  Code    &  13.00\%  & 0.055 & 872~GB \\
  Natural language related to code & 1.80\%  & 0.001 & 942~GB \\
  Natural language    & 0.20\%  & $6.9\mathrm{e}{-5}$ & 938~GB \\
  \midrule
\multicolumn{4}{l}{\textbf{Flag tuning fine-tuning (additional 84 billion tokens)}} \\ 
  Flag tuning & 85.00\%  & 1.700 & 103~GB\\ 
  Compiler emulation & 11.73\%  & 0.136 & 175~GB \\ 
  Code    & 2.84\%  & 0.007 & 872~GB \\
  Natural language related to code & 0.40\%  & $1.1\mathrm{e}{-4}$  & 942~GB \\
  Natural language    & 0.03\%  & $8.8\mathrm{e}{-6}$ & 938~GB \\
  \midrule 
\multicolumn{4}{l}{\textbf{Disassembly fine-tuning (additional 80 billion tokens)}} \\ 
  Disassembly & 85.00\%  & 1.707 & 88~GB \\
  Flag tuning & 4.68\%  & 0.089 & 103~GB \\ 
  Compiler emulation & 8.07\%  & 0.089 & 175~GB \\ 
  Code    & 1.96\%  & 0.004 & 872~GB \\
  Natural language related to code & 0.27\%  & $7.5\mathrm{e}{-5}$ & 942~GB \\
  Natural language    & 0.03\%  & $5.7\mathrm{e}{-6}$ & 938~GB \\
  \bottomrule  
    \end{tabular}    
    \caption{Training datasets used.
    \label{tab:dataset}}
\end{table}

\subsection{Pretraining on assembly code and compiler IRs}
\label{sec:pretraining}

The data used to train coding LLMs are typically composed largely of high level source languages like Python. Assembly code contributes a negligible proportion of these datasets, and compiler IRs even less. To build an LLM with a good understanding of these languages we initialize \foundationmodel models with the weights of \codellama and then train for 401 billion tokens on a compiler-centric dataset composed mostly of assembly code and compiler IRs, shown in Table~\ref{tab:dataset}.

\paragraph{Dataset}

\foundationmodel is trained predominantly on compiler intermediate representations and assembly code generated by LLVM~\citep{llvm} version 17.0.6. These are derived from the same dataset of publicly available code used to train \codellama. We summarize this dataset in Table~\ref{tab:dataset-stats}. As in \codellama, we also source a small proportion of training batches from natural language datasets.

\begin{table}
\small
\begin{subtable}[t]{0.43\textwidth}
\centering
\begin{tabular}{llll}
\toprule
 &   Items &   Tokens & Disk size \\
\midrule
LLVM-IR  &  10.7 M &  185 B &  432 GB \\
Assembly &  10.1 M &  216 B &  440 GB \\
\midrule
Total & 20.8 M & 401 B & 872 GB\\
\bottomrule
\end{tabular}
\caption{Language}
\end{subtable}
\hspace{\fill}
\begin{subtable}[t]{0.55\textwidth}
\centering
\begin{tabular}{llll}
\toprule
{} &   Items &   Tokens & Disk size \\
\midrule
x86\_64-unknown-linux-gnu  &  17.3~M &  340.3~B &    738~GB \\
aarch64-unknown-linux-gnu &   3.5~M &  60.5~B &  133~GB \\
nvptx64-nvidia-cuda       &   9.2~k &    146~M &  286~MB \\
\midrule
Total & 20.8 M & 401 B & 872 GB\\
\bottomrule
\end{tabular}
\caption{Target}
\end{subtable}
\caption{Composition of data used for initial IR and assembly pretraining. \foundationmodel is trained on a near-even split of IR and assembly code, predominantly targeting x86-64 architecture, with some 64-bit ARM, and a small amount of CUDA.}
\label{tab:dataset-stats}
\end{table}

\subsection{Instruction fine-tuning for compiler emulation}
\label{sec:code_change}

\begin{figure}
\begin{center}
\centerline{\includegraphics[width=\textwidth]{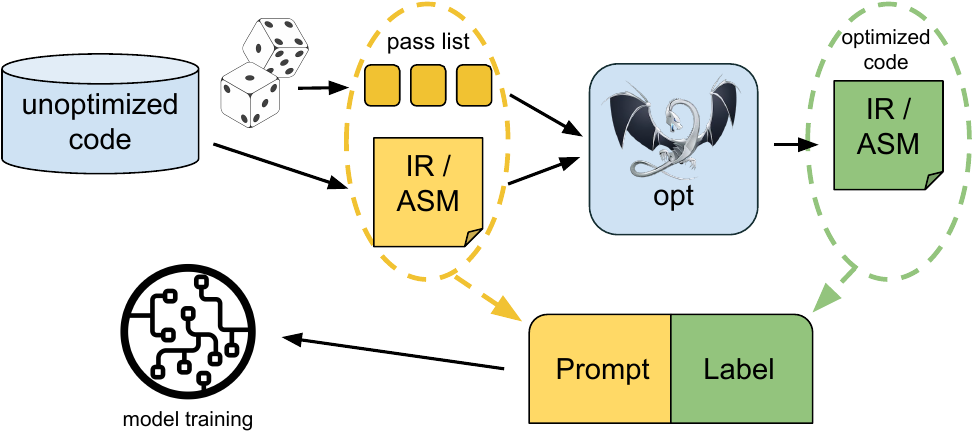}}
\caption{%
    To give the model an understanding of how compiler optimizations work, we use \emph{compiler emulation}. Unoptimized code samples and random pass lists are given to \lstinline{opt} to generate optimized code (IR or assembly). Pass list and input code are taken together as prompt while the generated output code is used as label.
}
\label{fig:code_change}
\end{center}
\end{figure}

To understand the mechanism of code optimization we instruction fine-tune \foundationmodel models to emulate compiler optimizations, illustrated in Figure~\ref{fig:code_change}. The idea is to generate from a finite set of unoptimized seed programs a large number of examples by applying randomly generated sequences of compiler optimizations to these programs. We then train the model to predict the code generated by the optimizations. We also train the model to predict the code size after the optimizations have been applied.

\paragraph{Task specification.} Given unoptimized LLVM-IR (as emitted by the \emph{clang} frontend), a list of optimization passes, and a starting code size, generate the resulting code after those optimizations have been applied and the resulting code size.

There are two flavors of this task: in the first the model is expected to output compiler IR, in the second the model is expected to output assembly code. The input IR, optimization passes, and code size are the same for both flavors. The prompt dictates the required output format. Examples of each prompt are provided in Appendices Listings~\ref{fig:compiler-emulation-prompt-ir} and~\ref{fig:compiler-emulation-prompt-asm}.

\paragraph{Code size.} We use two metrics for code size: the number of IR instructions, and \emph{binary size}. Binary size is computed by summing the size of the \texttt{.TEXT} and \texttt{.DATA} sections of the IR or assembly after lowering to an object file; we exclude \texttt{.BSS} section from our binary size metric since it does not affect on-disk size.

\paragraph{Optimization passes.} In this work we target LLVM 17.0.6 and use the New Pass Manager~\citep{llvm_newpassmanager} which classifies passes for different levels such as \textit{module}, \textit{function}, \textit{loop}, etc. as well as transformation and analysis passes. Transformation passes change given input IR while analysis passes generate information that influence subsequent transformations.

\begin{figure}
\setcaptiontype{lstlisting}
\small
\begin{lstlisting}
opt input.bc -o output.bc -p 'module(default<Oz>),module(iroutliner)'
clang output.bc -o output.o
size output.o
\end{lstlisting}
\caption{Commands used to apply an optimization pipeline comprising -Oz passes followed by IR outlining to an unoptimized IR \texttt{input.bc}. Binary size is the sum of \texttt{.TEXT} and \texttt{.DATA} section sizes of the lowered object file as reported by \texttt{size}.
\label{lst:flag-tuning-commands}}
\end{figure}

Of the 346 possible pass arguments for \emph{opt}, we select 167 to use. This includes each of the default optimization pipelines (e.g. \texttt{module(default<Oz>)}), individual optimization transform passes (e.g. \texttt{module(constmerge)}), but excludes non-optimization utility passes (e.g. \texttt{module(dot-callgraph)}) and transformations passes that are not semantics preserving (e.g. \texttt{module(internalize)}). We exclude analysis passes since they have no side effects and we rely on the pass manager to inject dependent analysis passes as needed.
For passes that accept parameter arguments we use the default values (e.g. \texttt{module(licm<allowspeculation>)}). Table~\ref{tab:passes} contains a list of all passes used. We used LLVM's \emph{opt} tool to apply pass lists and \emph{clang} to lower the resulting IR to object file. Listing~\ref{lst:flag-tuning-commands} shows the commands used.

\paragraph{Dataset.} We generated the compiler emulation dataset by applying random lists of between 1 and 50 optimization passes to unoptimized programs summarized in Table~\ref{tab:dataset-stats}. The length of each pass list was selected uniformly at random. Pass lists were generated by uniformly sampling from the set of 167 passes described above. Pass lists which resulted in compiler crashes or timed out after 120 seconds were excluded.

\section{\model: Extending for downstream compiler tasks}
\label{sec:downstream-method}

\subsection{Instruction fine-tuning for optimization flag tuning}
\label{sec:flag-tuning}

Manipulating compiler flags is well known to have a considerable impact on both runtime performance and code size~\citep{fursin_evaluating_2005}. We train \model models on the downstream task of selecting flags for LLVM's IR optimization tool \emph{opt} to produce the smallest code size.  Machine learning approaches to flag tuning have shown good results previously, but struggle with generalizing across different programs~\citep{compilergym}. Previous works usually need to compile new programs tens or hundreds of times to try out different configurations and find out the best-performing option. We train and evaluate \model models on the zero-shot version of this task by predicting flags to minimize code size of unseen programs. Our approach is agnostic to the chosen compiler and optimization metric, and we intend to target runtime performance in the future. For now, optimizing for code size simplifies the collection of training data.

\paragraph{Task specification.} We present the \model models with an unoptimized LLVM-IR (as emitted by the \emph{clang} frontend) and ask it to produce a list of \emph{opt} flags that should be applied to it, the binary size before and after these optimizations are applied, and the output code. If no improvement can be made over the input code, a short output message is generated that contains only the unoptimized binary size. Listings~\ref{fig:flag-tuning-prompt} and~\ref{fig:flag-tuning-prompt-no-improvement} provide the prompt and output templates for this task.

We used the same constrained set of optimization passes as in the compiler emulation task, and compute binary size in the same manner.

\begin{figure}
\begin{center}
\centerline{\includegraphics[width=1\textwidth]{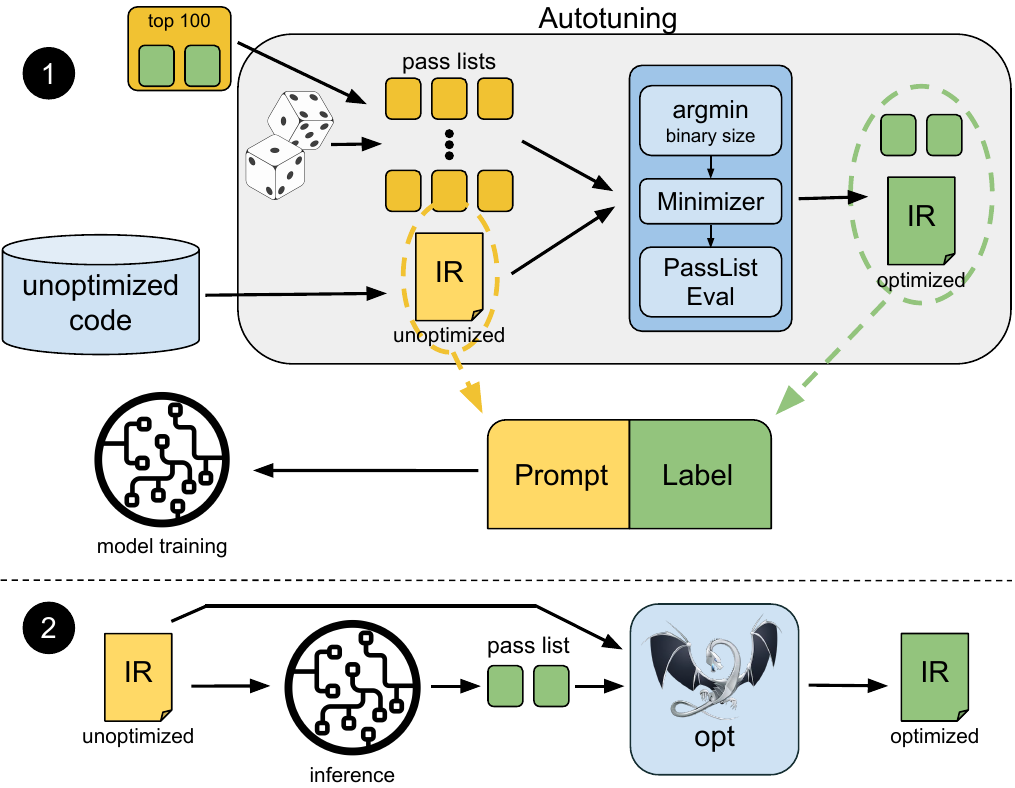}}
\caption{%
    Overview of our approach, showing the model input (Prompt) and output (Label) during training \textcircled{1} and inference \textcircled{2}. The prompt contains unoptimized code. The label contains an optimization pass list, binary size, and the optimized code. To generate the label for the training prompt, the unoptimized code is compiled against multiple random pass lists. The pass list achieving the minimum binary size is selected, minimized and checked for correctness with PassListEval. The final pass list together with its corresponding optimized IR are used as label during training. In a last step, the top 100 most often selected pass lists are broadcast among all programs. For deployment we generate only the optimization pass list which we feed into the compiler, ensuring that the optimized code is correct.%
    \label{fig:flag-tuning-overview}
}
\end{center}
\end{figure}

Figure~\ref{fig:flag-tuning-overview} illustrates the process used to generate training data (described below) and how the model is used for inference. Only the generated pass list is needed at evaluation time. We extract the pass list from the model output and run \textit{opt} using the given arguments. We can then evaluate the accuracy of the model predicted binary sizes and optimized output code, but those are auxiliary learning tasks not required for use.

\begin{figure}
\begin{center}
\centerline{\includegraphics[width=1\textwidth]{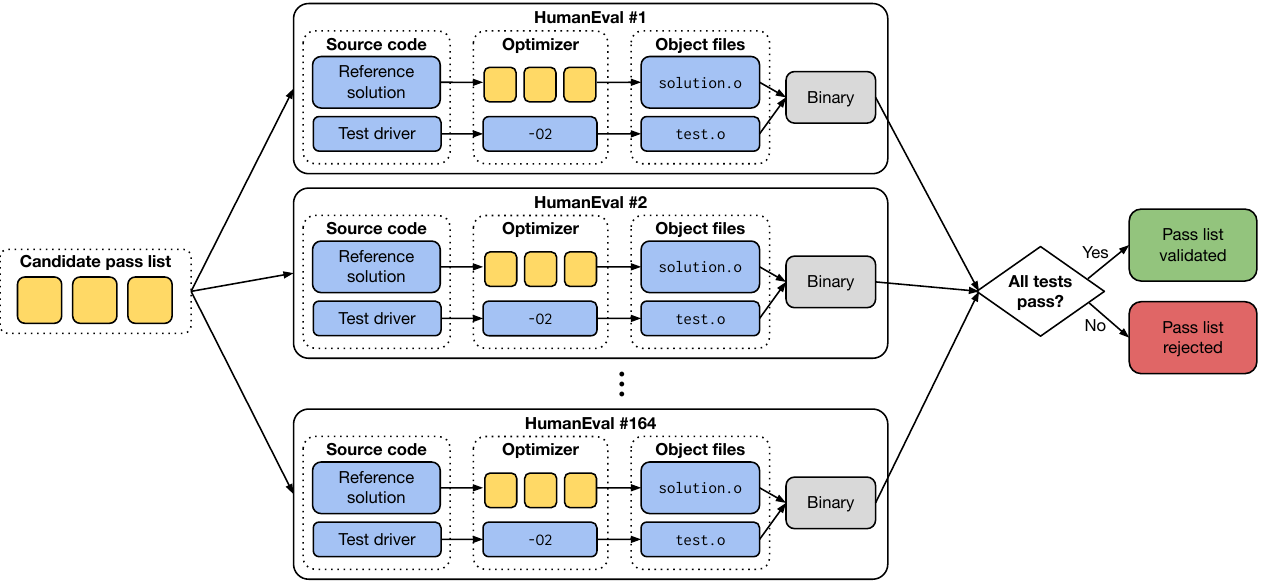}}
\caption{%
    Validating a candidate list of optimization passes using PassListEval. The candidate pass list is applied to the reference solutions for all 164 programs in HumanEval-X. The unit tests for these reference solutions are optimized using a conservative -O2 pass pipeline to ensure correctness, and then linked against the reference solutions. The resulting binaries are executed and if any of the binaries crash during execution, or if any of the compiler invocations fail, the pass list is rejected.  
}
\label{fig:passlisteval}
\end{center}
\end{figure}

\paragraph{Correctness.} LLVM's optimizer is not free from bugs and running optimization passes in unexpected or untested orders may expose subtle correctness errors that undermine the utility of the model. To mitigate this risk we developed \emph{PassListEval}, a tool to help in automatically identifying pass lists that break program semantics or cause compiler crashes. An overview of the tool is shown in Figure~\ref{fig:passlisteval}. PassListEval accepts as input a candidate pass list and evaluates it over a suite of 164 self-testing C++ programs, taken from HumanEval-X~\citep{humaneval-x}. Each program contains a reference solution for a programming challenge, e.g. \emph{``Check if in given vector of numbers, are any two numbers closer to each other than given threshold''}, and a suite of unit tests that validate correctness. We apply the candidate pass lists to the reference solution, and then link them against the test suites to produce a binary. When executed, the binary will crash if any of the tests fail. If any binary crashes, or if any of the compiler invocations fail, we reject the candidate pass list.

\paragraph{Dataset.} We trained \model models on a dataset of flag tuning examples derived from 4.5M of the unoptimized IRs used for pretraining. To generate the example optimal pass list for each program we ran an extensive iterative compilation process depicted in Figure~\ref{fig:flag-tuning-overview} and outlined below:

\begin{enumerate}
    \item We used large-scale \emph{random search} to generate an initial candidate best pass list for the programs. For each program we independently generated random lists of up to 50 passes by uniformly sampling from the set of 167 searchable passes described previously. Every time we evaluated a pass list on a program we recorded the resulting binary size. We then pick the per-program pass lists that produced the lowest binary size. We ran 22 billion unique compilations for an average 4,877 per program.
    \item The pass lists generated by random search may contain redundant passes that have no effect on the final outcome. Further, some pass orderings are commutative such that reordering then does not affect the final outcome. Since these would introduce noise in our training data, we developed a \emph{minimization} process which we applied to each pass list. Minimization comprises three steps: redundant pass elimination, bubble sort, and insertion search. In redundant pass elimination we minimize the best pass list by iteratively removing individual passes to see if they contribute to the binary size. If not, they are discarded. This is repeated until no further passes can be discarded. Bubble sort then attempts to provide a uniform ordering for pass subsequences by sorting passes based on a key. Finally, insertion sort performs a local search by iterating over each pass in the pass list and attempting to insert each of the 167 search passes before it. If doing so improves the binary size, this new pass list is kept. The entire minimization pipeline loops until a fixed point is reached. The distribution of minimized pass list lengths is shown in Figure~\ref{fig:pass-list-len}. The average pass list length is 3.84.
    \item We apply \emph{PassListEval}, described previously, to the candidate best pass lists. Through this we identified 167,971 of 1,704,443 unique pass lists (9.85\%) as causing compile time or runtime errors.
    \item We broadcast the \emph{top 100} most frequently optimal pass lists across all programs, updating the per-program best pass lists if improvements are found. After this the total number of unique best pass lists decreases from 1,536,472 to 581,076.
\end{enumerate}

The autotuning pipeline outlined above produced a geometric mean 7.1\% reduction in binary size over -Oz. Figure~\ref{fig:autotuner-best-pass-freq} shows the frequency of individual passes. For our purposes, this autotuning serves as a gold standard for the optimization of each program. While the binary size savings discovered are significant, this required 28 billion additional compilations at a computational cost of over 21,000 CPU days. The goal of instruction fine-tuning \model to perform the flag tuning task is to achieve some fraction of the performance of the autotuner without requiring running the compiler thousands of times.

\newpage

\subsection{Instruction fine-tuning for disassembly}
\label{sec:disassembly}

\begin{figure}
\begin{center}
\centerline{\includegraphics[width=1\textwidth]{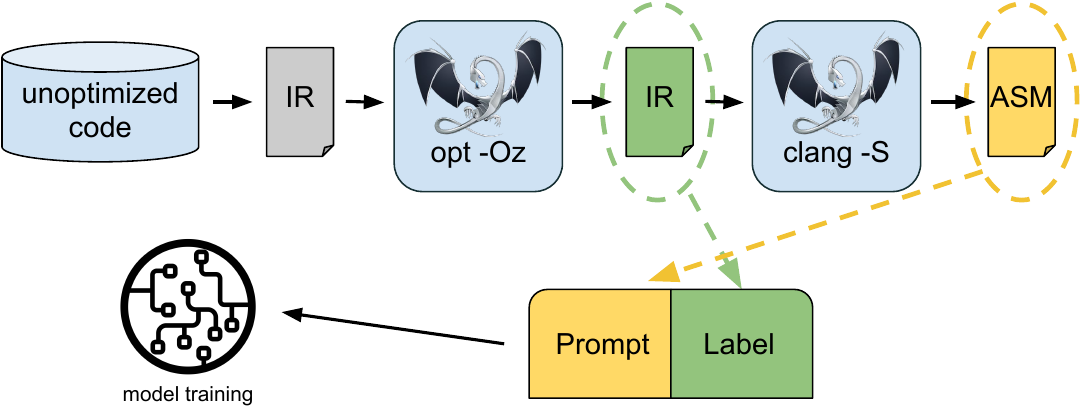}}
\caption{%
    We train the model to understand the relationship between assembly and IR by training it to disassemble a given code sample to its corresponding IR. The IR used to label this training task was generated by optimizing an IR with the -Oz flag.
    \label{fig:disassembly}
}
\end{center}
\end{figure}

The ability to lift code from assembly back into higher level structures 
enables running additional optimizations on library code directly integrated with application code or porting of legacy code to new architectures. The field of decompilation has seen advancements in applying machine learning techniques to generate readable and accurate code from binary executables. Several studies explore the use of machine learning for decompilation tasks, such as lifting binaries into intermediate representations for evaluation against synthetic C programs~\citep{cao2022boosting}, utilizing evolutionary approaches like genetic algorithms for program analysis~\citep{schulte2018evolving}, and proposing methods like XLIR for matching binary code across different programming languages~\citep{gui2022cross}. \citet{armengol2023slade} have trained a language model to decompile x86 assembly into high level C code. In this study, we demonstrate how \model can learn the relationship between assembly code and compiler IR by fine-tuning it for disassembly. The task is to learn the inverse translation of \texttt{clang -xir - -o - -S}, shown in Figure~\ref{fig:disassembly}.

\paragraph{Round tripping.} Using an LLM for disassembly causes problems of correctness. The lifted code must be verified by an equivalence checker which is not always feasible or manually verified for correctness or subjected to sufficient test cases to give confidence. However, a lower bound on correctness can be found by round-tripping. That is to say by compiling the lifted IR back into assembly, if the assembly is identical then the IR is correct. This gives an easy route to using the results of the LLM and an easy way to measure the utility of a disassembly model.

\paragraph{Task specification.} We provide the model with assembly code and train it to emit the corresponding disassembled IR. Listing~\ref{fig:disassembly-prompt} shows the prompt format. The context length for this task is set to 8k tokens for the input assembly code and 8k tokens for the output IR.

\paragraph{Dataset.} We derive the assembly codes and IR pairs from the same dataset used in previous tasks. Our fine-tuning dataset consists in 4.7M samples. The input IR has been optimized with -Oz before being lowered to x86 assembly.

\section{Training parameters}

Data is tokenized via byte pair encoding~\citep{bpe}, employing the same tokenizer as \codellama, Llama~\citep{llama1}, and Llama 2~\citep{llama}.

We use the same training parameters for all four stages of training. Most of the training parameters we used are the same as for the \codellama base model. We use the AdamW \citep{adamw} optimizer with $\beta_1$ and $\beta_2$ values of 0.9 and 0.95. We use a cosine schedule with $1000$ warm-up steps, and set the final learning rate to be 1/30th of the peak learning rate. Compared to the \codellama base model, we increased the context length of individual sequences from 4,096 to 16,384, but kept the batch size constant at 4M tokens. To account for the longer context, we set our learning rate to $2e^{-5}$ and modified the parameters of the \textsc{RoPE} positional embeddings~\citep{su2024roformer} where we reset frequencies with a base value of $\theta = 10^{6}$. These settings are in accordance with the long context training done for the \codellama base model.
\section{Evaluation}
\label{sec:results}

In this section we evaluate the performance of \foundationmodel models on the tasks of flag tuning and disassembly, compiler emulation, next-token prediction, and finally software engineering tasks.

\subsection{Flag tuning task}
\label{sec:flag_tuning_results}

\paragraph{Methodology.} We evaluate \model on the task of optimization flag tuning for unseen programs and compare to GPT-4 Turbo and \codellamainst. We run inference on each model and extract from the model output the optimization pass list. We then use this pass list to optimize the particular program and record the binary size. The baseline is the binary size of the program when optimized using -Oz.

For GPT-4 Turbo and \codellamainst we append a suffix to the prompt with additional context to further describe the problem and expected output format. After some experimentation we found that the prompt suffix shown in Listing~\ref{fig:flag-tuning-prompt-suffix} provides the best performance.

All model-generated pass lists are validated using \emph{PassListEval}, and -Oz is used as substitute if validation fails. To further validate correctness of model-generated pass lists we link the final program binaries and differential test their outputs against the outputs of the benchmark when optimized using a conservative -O2 optimization pipeline.

\paragraph{Dataset.} We evaluate on 2,398 test prompts extracted from the MiBench benchmark suite~\citep{mibench}. To generate these prompts we take all of the 713 translation units that make up the 24 MiBench benchmarks and generate unoptimized IRs from each. We then format them as prompts as per Listing~\ref{fig:flag-tuning-prompt}. If the resulting prompt exceeds 15k tokens we split the LLVM module representing that translation unit into smaller modules, one for each function, using \emph{llvm-extract}. This results in 1,985 prompts which fit within the 15k token context window, leaving 443 translation units which do not fit. We use -Oz when for the 443 excluded translation units when computing performance scores. Table~\ref{tab:mibench-overview} summarizes the benchmarks.

\paragraph{Results.} Table~\ref{tab:flag-tuning-perf} shows zero-shot performance of all models on the flag tuning task. Only \model models provide an improvement over -Oz, with the 13B parameter model marginally outperforming the smaller model, generating smaller object files than -Oz in 61\% of cases.

\begin{figure}
\begin{center}
\centerline{\includegraphics[width=1\textwidth]{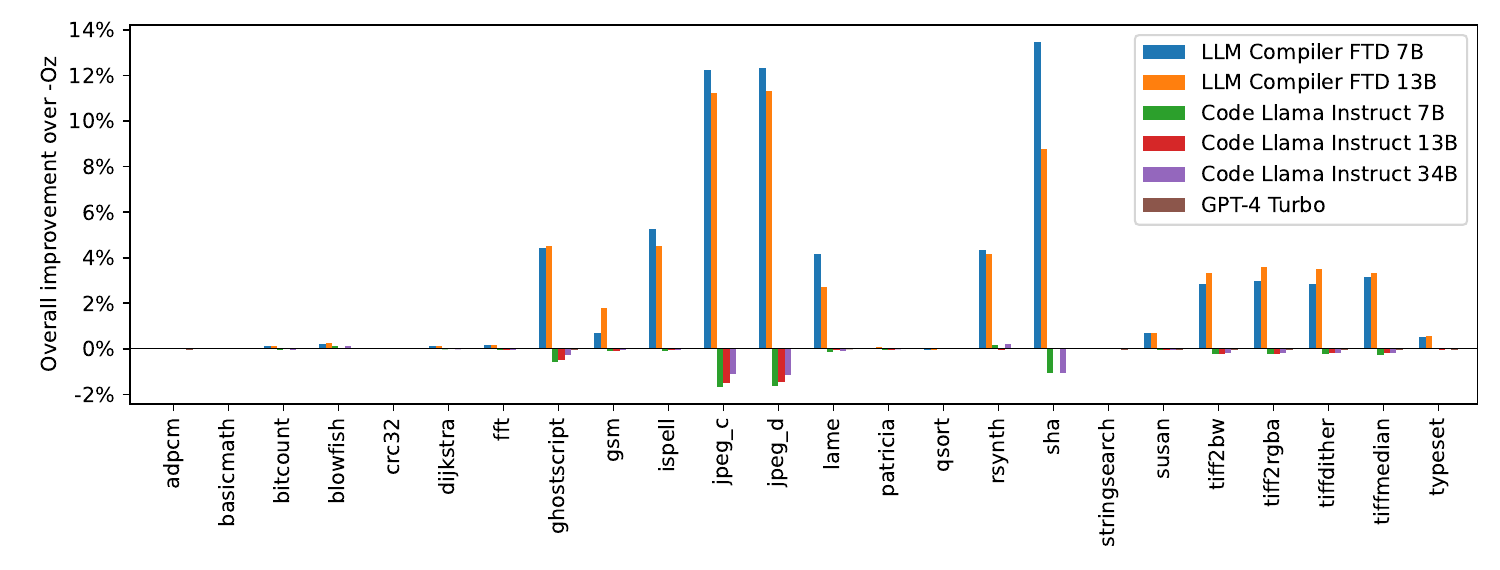}}
\caption{%
    Improvement over -Oz for each of the benchmarks in MiBench. 
}
\label{fig:flag-tuning-perf-by-benchmark}
\end{center}
\end{figure}

In some cases the model-generated pass list causes a larger object file size than -Oz. For example, \model 13B regresses in 12\% of cases. These regressions can be avoided by simply compiling the program twice: once using the model-generated pass list, once using -Oz, and selecting the pass list which produces the best result. By eliminating regressions wrt -Oz, these \emph{-Oz backup} scores raise the overall improvement over -Oz to 5.26\% for \model 13B, and enable modest improvements over -Oz for \codellamainst and GPT-4 Turbo. Figure~\ref{fig:flag-tuning-perf-by-benchmark} shows the performance of each model broken down by individual benchmark.

\begin{figure}
\begin{center}
\begin{subfigure}{\textwidth}
\centering
\includegraphics[width=\textwidth]{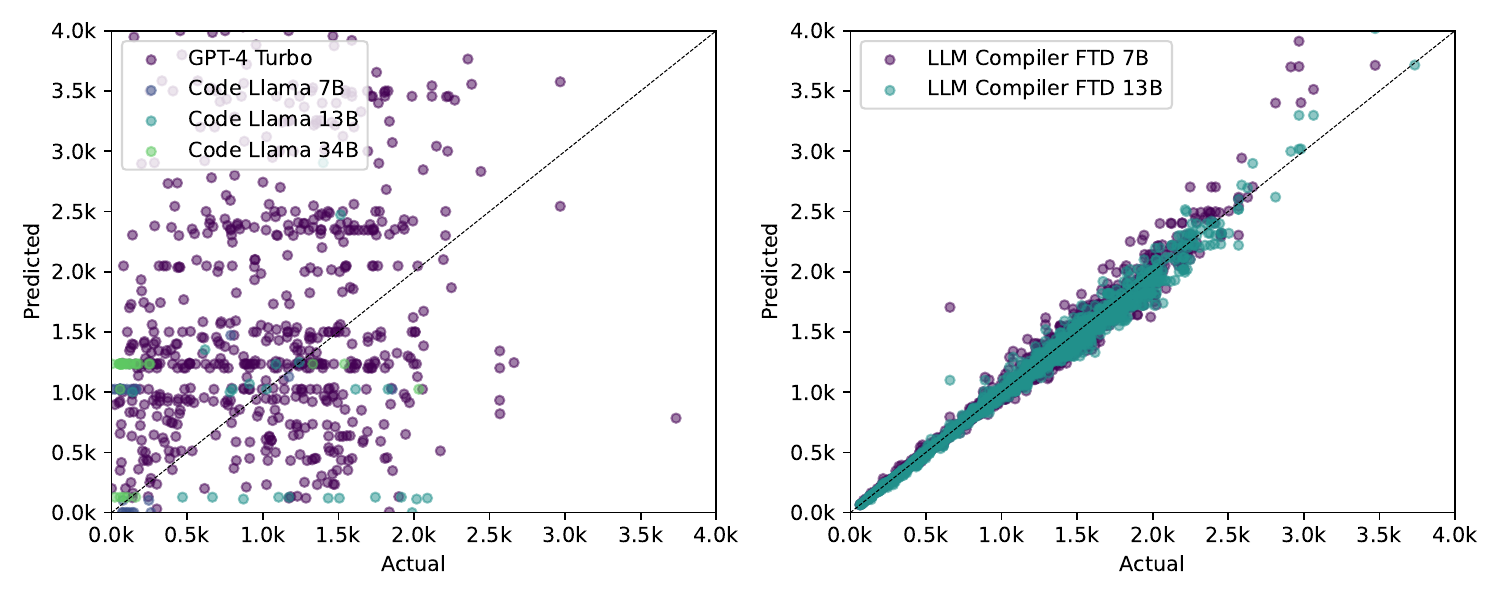}
\caption{Unoptimized binary size}
\label{fig:sub1}
\end{subfigure}
\begin{subfigure}{\textwidth}
\centering
\includegraphics[width=\textwidth]{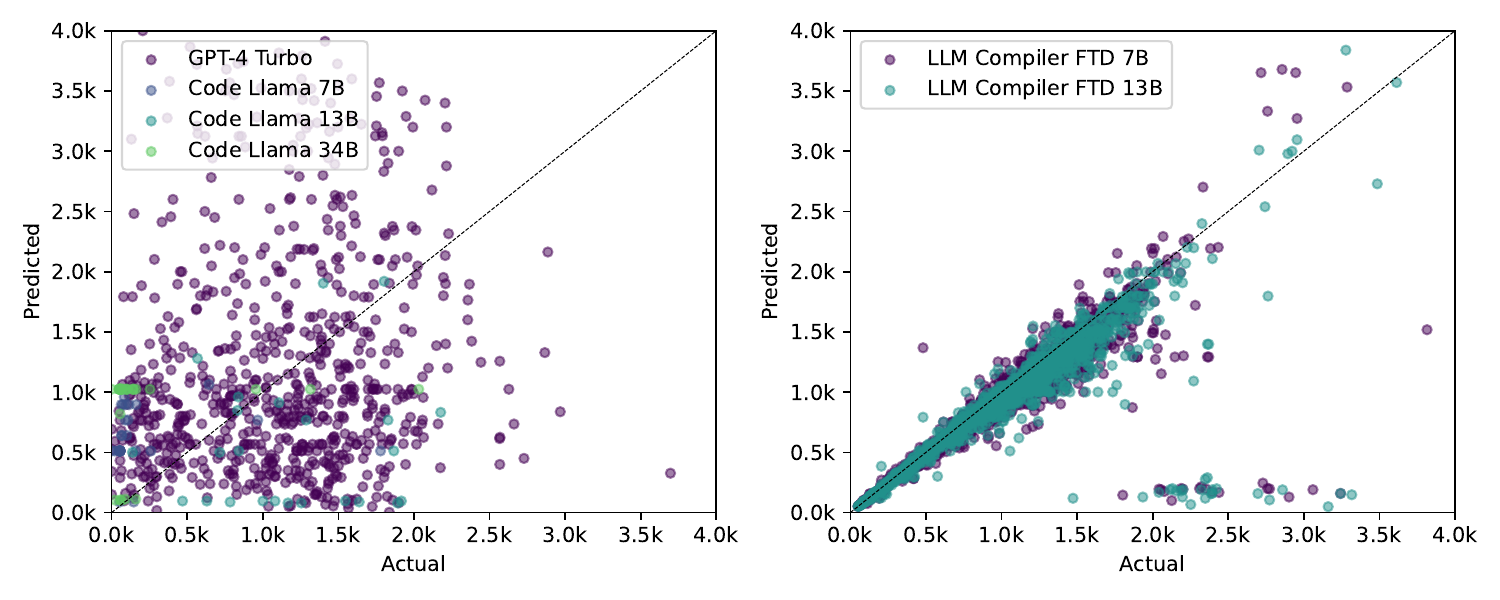}
\caption{Optimized binary size}
\label{fig:sub2}
\end{subfigure}
\caption{%
    Accuracy of models at predicting code size before and after optimization. \model is most accurate at predicting code size before optimization than after optimization. \codellama and GPT-4 Turbo, shown left, display little correlation between predicted and actual values.
}
\label{fig:binary-size-predictions}
\end{center}
\end{figure}

\paragraph{Binary size accuracy.} While the model-generated binary size predictions have no effect on actual compilation, we can evaluate the performance of the models at predicting binary sizes before and after optimization to give an indication of each model's understanding of optimization. Figure~\ref{fig:binary-size-predictions} shows the results. \model binary size predictions correlate well with ground truth, with the 7B parameter model achieving MAPE values of 0.083 and 0.225 for unoptimized and optimized binary sizes respectively. The 13B parameter model improved has similar MAPE values of 0.082 and 0.225. \codellamainst and GPT-4 Turbo binary size predictions show little correlation with ground truth. We note that the \model errors are slightly higher for optimized code than unoptimized code. In particular, there is an occasional tendency for \model to overestimate the effectiveness of optimization, resulting in a lower predicted binary size than actual.

\begin{table}
\centering
\small
\caption{Comparison of model performance when flag tuning 2,398 object files from MiBench. \emph{Overall improvement} scores include 443 object files which do not fit in the context window of \model. For GPT-4 and the \codellama models we appended a suffix to the prompt to provide additional context (see Listing~\ref{fig:flag-tuning-prompt-suffix} in the Appendix). %
\label{tab:flag-tuning-perf}}
\vspace{.5em}
\begin{tabular}{lr|C{1.5cm}C{1.5cm}|cc}
\toprule
&&&&\multicolumn{2}{c}{Overall improvement over -Oz}\\
& Size & Improved & Regressed & zero-shot & -Oz backup \\
\midrule
\multirow{2}{*}{\model} & 7B & 1,465 & 302 & 4.77\% & 5.24\% \\
& 13B & \textbf{1,466} & \textbf{299} & \textbf{4.88\%} & \textbf{5.26\%} \\
\midrule
\multirow{3}{*}{\codellamainst} & 7B & 379 & 892 & -0.49\% & 0.23\% \\
& 13B & 319 & 764 & -0.42\% & 0.18\% \\
& 34B & 230 & 493 & -0.27\% & 0.15\% \\
\midrule
GPT-4 Turbo (2024-04-09) & - & 13 & 24 & -0.01\% & 0.03\% \\
\bottomrule
\end{tabular}
\end{table}

\paragraph{Ablation studies.}

Table~\ref{tab:autotuning-ablation} ablates the performance of models on a small holdout validation set of 500 prompts taken from the same distribution as our training data (though not used during training). We trained for flag tuning at each stage of the training pipeline from Figure~\ref{fig:training-pipeline} to compare performance. As shown, disassembly training causes a slight regression in performance from average 5.15\% to 5.12\% improvement over -Oz. We also show performance of the autotuner used for generating the training data described in Section~\ref{sec:method}. \model achieves 77\% of the performance of the autotuner.

\begin{table}
\centering
\footnotesize
\caption{%
  Ablating the \model training regime on the flag tuning task. All results are for 7B parameter models, evaluated on the same holdout validation set of 500 programs. The first row is the \model release. All other rows strip out successful components of the training regime. %
  \label{tab:autotuning-ablation}
}
\vspace{.5em}
\begin{tabular}{C{1.8cm}C{1.5cm}C{2cm}C{1.8cm}C{1.8cm}|ll}
\toprule
&&&&&\multicolumn{2}{c}{Mean improvement}\\
\codellama & IR \& asm pretraining & Compiler emulation & Flag tuning & Disassembly & over -Oz & wrt. Autotuner \\
\midrule
\cmark & \cmark & \cmark & \cmark & \cmark & 5.12\% & 77\% \\
\cmark & \cmark & \cmark & \cmark & & \textbf{5.15\%} & \textbf{78\%} \\
\cmark & \cmark & & \cmark & & 5.07\% & 76\% \\
\cmark & & & \cmark & & 4.94\% & 75\% \\
 & & & \cmark & & 4.79\% & 72\% \\
\midrule
\multicolumn{5}{l}{Autotuner} & 6.63\% & 100\%\\
\bottomrule
\end{tabular}
\end{table}

\subsection{Disassembly task}

\paragraph{Methodology.} We evaluate the functional correctness of LLM-generated code when disassembling assembly code to LLVM-IR. As in Section~\ref{sec:flag_tuning_results} we evaluate \model and compare to \codellamainst and GPT-4 Turbo, and find that an additional prompt suffix, shown in Listing~\ref{fig:disassembly-prompt-suffix}, is required to extract the best performance from these models. The suffix provides additional context about the task and the expected output format. To evaluate the performance of models we \emph{round-trip} the model-generated disassembled IR back down to assembly. This enables us to evaluate accuracy of the disassembly by comparing the BLEU score~\citep{bleu} of the original assembly against the round-trip result. A lossless and perfect disassembly from assembly to IR will have a round-trip BLEU score of 1.0 (\emph{exact match}).

\paragraph{Dataset.} We evaluate on 2,015 test prompts extracted from the MiBench benchmark suite. We took the 2,398 translation units used for the flag tuning evaluation above and generated disassembly prompts. We then filtered the prompts on a maximum 8k token length, allowing 8k tokens for the model output, leaving 2,015. Table~\ref{tab:mibench-disassembly-overview} summarizes the benchmarks.

\begin{figure}
\begin{center}
\centerline{\includegraphics[width=.5\textwidth]{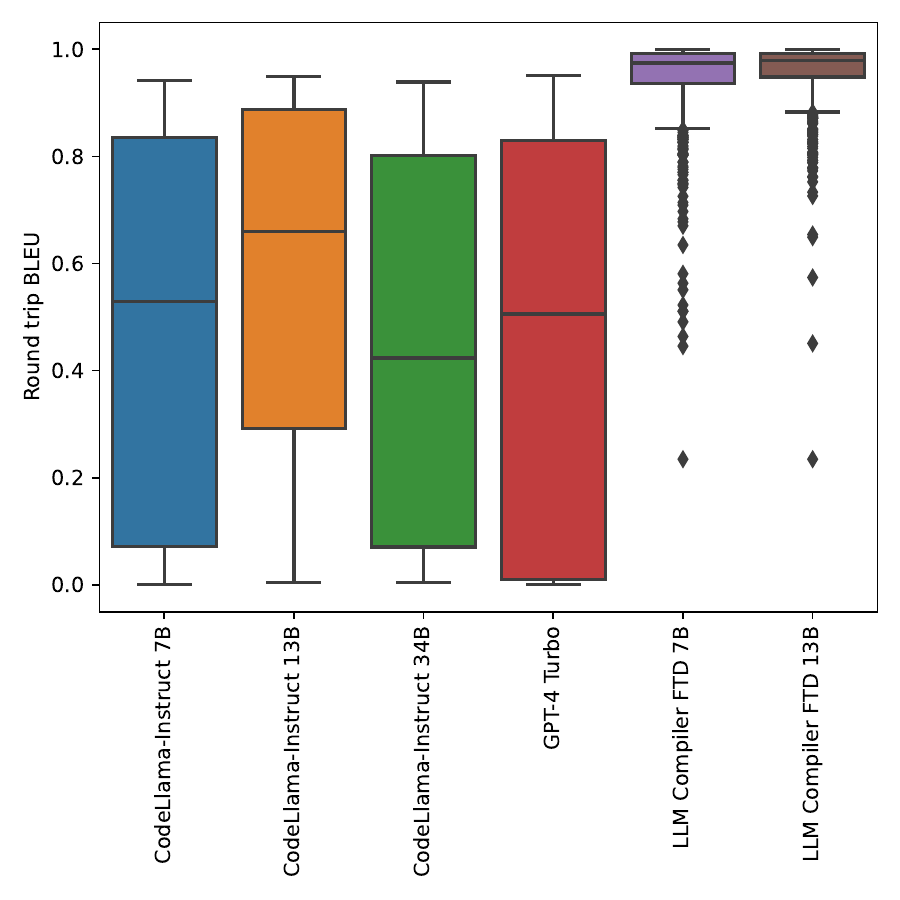}}
\caption{%
    Distribution of round trip BLEU scores on the disassembly task.
}
\label{fig:disassembly-bleu-boxplot}
\end{center}
\end{figure}

\paragraph{Results.} Table~\ref{tab:disassembly-perf} shows performance of the models on the disassembly task. \model 7B has a slightly higher round-trip success rate than \model 13B, but \model 13B has the highest accuracy of round-tripped assembly (\emph{round trip BLEU}) and most frequently produces a perfect disassembly (\emph{round trip exact match}). \codellamainst and GPT-4 Turbo struggle with generating syntactically correct LLVM-IR. Figure~\ref{fig:disassembly-bleu-boxplot} shows the distribution of round-trip BLEU scores for all models.

\begin{table}
\centering
\small
\caption{Model performance at disassembling 2,015 assembly codes taken from MiBench. We use \emph{Round trips} to evaluate the capabilities of models, by taking the IR generated by the models and attempting to lower it back to assembly. \emph{Round trips} shows the number of disassembled IRs that can be lowered back, \emph{Round trip BLEU} compares the round-tripped assemblies against the originals, and \emph{Round trip exact match} is the proportion of round-tripped assemblies that are exact character-for-character matches with the input, indicating lossless round-trip from assembly up to IR and back down again.}
\vspace{.5em}
\begin{tabular}{lr|ccc}
\toprule
& Size & Round trips & Round trip BLEU & Round trip exact match \\
\midrule
\multirow{2}{*}{\model} & 7B & \textbf{936} & 0.951 & 12.7\% \\
& 13B &  905 & \textbf{0.960} & \textbf{13.8\%} \\
\midrule
\multirow{3}{*}{\codellamainst} & 7B & 30 & 0.477 & 0.0\%\\
& 13B & 53 & 0.615 & 0.0\% \\
& 34B & 12 & 0.458 & 0.0\% \\
\midrule
GPT-4 Turbo (2024-04-09) & - & 127 & 0.429 & 0.0\% \\
\bottomrule
\label{tab:disassembly-perf}
\end{tabular}
\end{table}

\paragraph{Ablation studies.}

Table~\ref{tab:disassembly-ablation} ablates the performance of models on a small holdout validation set of 500 prompts taken from the MiBench dataset used previously. We trained for disassembly at each stage of the training pipeline from Figure~\ref{fig:training-pipeline} to compare performance. Round trip rate is highest when going through the whole stack of training data and drops consistently with every training stage, though round trip BLEU varies little with each stage.

\begin{table}
\centering
\footnotesize
\caption{Ablating the \model training regime on code disassembly. All results are for 7B parameter model sizes, evaluated on a holdout validation set of 500 programs. Values in parentheses show relative performance to the first row (i.e. the \model release).}
\vspace{.5em}
\begin{tabular}{C{1.8cm}C{1.5cm}C{2cm}C{1.8cm}C{1.8cm}|L{2.3cm}L{2.3cm}}
\toprule
\codellama & IR \& asm pretraining & Compiler emulation & Flag tuning training & Disassembly &  Round trips & Round trip BLEU \\
\midrule
\cmark & \cmark & \cmark & \cmark & \cmark & \textbf{49.4\% (-)} & 0.951 (-) \\
\cmark & \cmark & \cmark & & \cmark & 45.2\% (-8.5\%) & 0.955 (+0.4\%) \\
\cmark & \cmark & & & \cmark & 44.2\% (-10.5\%) & 0.957 (+0.7\%) \\
\cmark & & & & \cmark & 39.0\% (-21.1\%) & \textbf{0.965 (+1.5\%)} \\
 & & & & \cmark & 8.8\% (-82.8\%) & 0.908 (-4.5\%) \\
\bottomrule
\label{tab:disassembly-ablation}
\end{tabular}
\end{table}

\subsection{Foundation model tasks}

\paragraph{Methodology} We ablate \foundationmodel models on the two foundation model tasks of next-token prediction and compiler emulation. We perform this evaluation at each stage of the training pipeline to see how training for each successive task affects performance. For next-token prediction we compute perplexity on a small sample of LLVM-IR and assembly code from all optimization levels. We evaluate compiler emulation using two metrics: whether the generated IR or assembly code compiles, and whether the generated IR or assembly code is an exact match for what the compiler would produce.

\paragraph{Dataset.} For next-token prediction we use a small holdout set of validation data that is drawn from the same distribution as our training data but has not been used for training. We use a mixture of optimization levels including unoptimized code, code optimized with -Oz, and randomly generated pass lists. For compiler emulatino we evaluate using 500 prompts generated from MiBench using randomly pass lists generated in the manner described in Section~\ref{sec:code_change}.

\paragraph{Results} Table~\ref{tab:compiler-eval-tasks} shows performance of \model across all training stages on the two foundation model training tasks of next-token prediction and compiler emulation. Next-token prediction performance jumps sharply after \codellama, which has seen very little IR and assembly, and declines slightly with each subsequent stage of fine-tuning.

For compiler emulation, the \codellama base model and the pre-trained models perform poorly since they have not been trained on this task. The highest performance is achieved directly after compiler emulation training where 95.6\% of IR and assembly generated by \model 13B compiles, and 20\% of it matches the compiler exactly. Performance declines after fine-tuning for flag tuning and disassembly.

\begin{table}
\scriptsize
\centering
\caption{Performance at next-token prediction and compiler emulation tasks. For \emph{Perplexity}, lower is better. For \emph{Compiles} and \emph{Exact match}, higher is better.}
\vspace{.5em}
\begin{tabular}{cC{1.5cm}C{1.7cm}C{1.4cm}C{1.5cm}r|cc|cc}
\toprule
&&&&&& \multicolumn{2}{c}{Perplexity} & \multicolumn{2}{c}{Compiler emulation}\\
\codellama & IR \& asm pretraining & Compiler emulation & Flag tuning & Disassembly & Size & IR & Asm & Compiles & Exact match \\
\midrule
\multirow{2}{*}{\cmark} & & & & & 7B & 1.456 & 1.423 & 5.4\% & 1.2\% \\
 & & & & & 13B & 1.429 & 1.404 & 4.8\% & 0.8\% \\
\midrule
\multirow{2}{*}{\cmark} & \multirow{2}{*}{\cmark} & & & & 7B & 1.050 & 1.041  & 0.8\% & 0.0\% \\
 & & & & & 13B & \textbf{1.045} & \textbf{1.038} & 35.8\% & 2.8\% \\
\midrule
\multirow{2}{*}{\cmark} & \multirow{2}{*}{\cmark} & \multirow{2}{*}{\cmark} & & & 7B & 1.052 & 1.046  & 87.0\% & 16.0\% \\
&  & & & & 13B & 1.047 & 1.043  & \textbf{95.6\%} & \textbf{20.0\%} \\
\midrule
\multirow{2}{*}{\cmark} & \multirow{2}{*}{\cmark} & \multirow{2}{*}{\cmark} & \multirow{2}{*}{\cmark} & & 7B & 1.058 & 1.051  & 55.0\% & 1.2\% \\
 &  & & & & 13B & 1.052 & 1.048  & 58.6\% & 4.2\% \\
\midrule
\multirow{2}{*}{\cmark} & \multirow{2}{*}{\cmark} & \multirow{2}{*}{\cmark} & \multirow{2}{*}{\cmark} & \multirow{2}{*}{\cmark} & 7B & 1.057 & 1.053 & 71.0\% & 4.6\%  \\
&  & & & & 13B & 1.054 & 1.052 & 61.4\% & 5.4\%  \\
\bottomrule
\label{tab:compiler-eval-tasks}
\end{tabular}
\end{table}

\subsection{Software engineering tasks}

\paragraph{Methodology.} While the purpose of \model is to provide foundation models for code optimization, it builds upon base \codellama models which were trained for software engineering tasks. To evaluate how the additional training of \model has affected the performance of code generation we use the same benchmark suites as in \codellama that evaluate the ability of LLMs to generate Python code from natural language prompts, such as \emph{``Write a function to find the longest chain which can be formed from the given set of pairs.''}.

\paragraph{Datasets.} We use the HumanEval~\citep{humaneval} and MBPP~\citep{mbpp} benchmarks as in \codellama.

\paragraph{Results.} Table~\ref{tab:human-eval-mbpp-scores} shows the greedy decoding performance (\emph{pass@1}) of all model training stages and model sizes starting at the \codellama base model. It also shows the models' scores on \emph{pass@10} and \emph{pass@100} which were generated with p=0.95 and temperature=0.6. Each stage of compiler-centric training causes a slight regression in Python programming ability. pass@1 performance on HumanEval and MBPP declines by up to 18\% and 5\% for \foundationmodel and by up to 29\% and 22\% for \model after the additional flag tuning and disassembly fine-tuning. All models still outperform Llama~2 on both tasks.

\begin{table}
\footnotesize
\centering
\caption{Performance on Python programming tasks. pass@1 are computed with greedy decoding. The pass@10 and pass@100 scores are computed with nucleus sampling with p=0.95 and temperature=0.6.}

\begin{tabular}{cccccr|ccc|ccc}
\toprule
\rotatebox[origin=lB]{90}{\codellama} & \rotatebox[origin=lB]{90}{\parbox{1.5cm}{IR \& asm pretraining}} & \rotatebox[origin=lB]{90}{\parbox{1.5cm}{Compiler emulation}} & \rotatebox[origin=lB]{90}{Flag tuning} & \rotatebox[origin=lB]{90}{Disassembly} & Size & \multicolumn{3}{c}{HumanEval} & \multicolumn{3}{c}{MBPP} \\
&&&&&&pass@1&pass@10&pass@100&pass@1&pass@10&pass@100\\
\midrule
\multirow{2}{*}{\cmark} & & & & & 7B & 32.9\% & 63.3\% & 85.3\% & 45.4\% & 67.5\% & 81.6\% \\
 & & & & & 13B & \textbf{36.0\%} & \textbf{71.9\%} & \textbf{90.6\%} & \textbf{48.4\%} & \textbf{71.3\%} & \textbf{83.9\%} \\
\midrule
\multirow{2}{*}{\cmark} & \multirow{3}{*}{\cmark} & & & & 7B & 28.0\% & 58.6\% & 84.3\% & 42.8\% & 66.0\% & 80.0\% \\
 &  & & & & 13B & 34.1\%  & 68.0\% & 87.9\% & 47.6\%  & 70.3\%  & 83.3\% \\
\midrule
\multirow{2}{*}{\cmark} & \multirow{3}{*}{\cmark} & \multirow{3}{*}{\cmark} & & & 7B & 25.0\% & 51.3\% & 79.0\%  & 37.4\% & 61.5\%  &  75.6\% \\
&  & & & & 13B & 31.1\% & 62.9\% & 83.2\% & 46.0\% &  67.8\%  &  80.9\% \\
\midrule
\multirow{2}{*}{\cmark} & \multirow{3}{*}{\cmark} & \multirow{3}{*}{\cmark} & \multirow{3}{*}{\cmark} & & 7B & 24.4\% & 46.2\% & 73.1\% & 36.6\% & 58.5\%  & 74.4\% \\
&  & & & & 13B & 29.3\% & 55.9\% & 81.1\% & 42.2\% & 63.6\%  & 79.1\% \\
\midrule
\multirow{2}{*}{\cmark} & \multirow{3}{*}{\cmark} & \multirow{3}{*}{\cmark} & \multirow{3}{*}{\cmark} & \multirow{3}{*}{\cmark} & 7B & 26.8\% & 44.0\% & 65.3\% & 31.4\% & 55.1\% & 73.2\% \\
&  & & & & 13B & 25.6\% & 51.2\% & 76.8\% & 37.6\% & 60.6\% & 76.4\% \\
\midrule
\multicolumn{5}{l}{\multirow{2}{*}{\textsc{Llama 2}}} & 7B & 12.2\% & 25.2\% & 44.4\%  & 20.8\% & 41.8\% & 65.5\% \\
\multicolumn{5}{l}{} & 13B & 20.1\% & 34.8\% & 61.2\% & 27.6\% & 48.1\% & 69.5\% \\
\bottomrule
\label{tab:human-eval-mbpp-scores}
\end{tabular}
\end{table}

\section{Related work}

\paragraph{Language models over code.} There is increasing interest in LLMs for source code reasoning and generation~\citep{jiang2024survey,llms-for-swe}. The main enablers of progress in this area are pretrained foundational models made available for others to build upon, including \codellama~\citep{codellama}, StarCoder~\citep{starcoder2}, Magicoder~\citep{wei2024magicoder}, DeepSeek-Coder~\citep{guo2024deepseekcoder}, GPT-4~\citep{openai2023gpt4} and others~\citep{wang2023codet5plus, allal2023santacoder, codebert}. Some of the existing models are open source~\citep{codellama, starcoder2, wei2024magicoder, allal2023santacoder} while others are closed source~\citep{humaneval, openai2023gpt4, li_competition-level_2022, gunasekar2023textbooks}. We extend the collection of foundational models for code with a family of models specifically trained on intermediate code representation with a license that allows wide reuse.

Language models have been adapted to perform program fuzzing~\citep{xia2023universal,titanfuzz}, test generation~\citep{schäfer2023adaptive}, automated program repair~\citep{apr_xia_2023}, and source-level algorithmic optimization \citet{madaan2023learning}. The introduction of fill-in-the-middle capabilities is especially useful for software engineering use cases such as code completion, and has become common in recent code models such as InCoder~\citep{fried_incoder_2023}, SantaCoder~\citep{allal2023santacoder}, StarCoder~\citep{starcoder2}, and \codellama~\citep{codellama}. A large number of useful applications have been explored for LLMs, however, only very few are directly focused on compilation tasks.

\paragraph{Language models over IR.} While LLMs have found broad adoption for coding tasks, few operate at the level of compilers. \citet{llvm-weakness-identification}~train a RoBERTA architecture on LLVM-IR for the purpose of code weakness identification, and Transcoder-IR~\citep{transcoder-ir} uses LLVM-IR as a pivot point for source-to-source translation. Few LLMs include compiler IRs in their training, and of those that do, IRs comprise a tiny fraction of the data compared to other programming languages. StarCoder~2~\cite{starcoder2} and DeepSeek-Coder~\cite{guo2024deepseekcoder} include 7.7~GB (0.4\%) and 0.91~GB (0.1\%) of LLVM-IR respectively in their training data. \foundationmodel is pretrained on 422~GB of LLVM-IR, and additional LLVM-IR during fine-tuning, and assembly code which makes up at least 85\% of the total training data.

\citet{paul2024ircoder} create SLTrans, a 26~B token dataset which pairs high level source code with corresponding LLVM-IR. Like our dataset, they include different source languages and optimization levels for their IR, however, their optimization is limited to \emph{-Oz} and \emph{-O3}. They train IRCoder on 800~M tokens of SLTrans and demonstrate how it improves the code reasoning capabilities of underlying base models. IRCoder and StarCoder~2 present their models with LLVM-IR. We include both LLVM-IR as well as native assembly code from multiple source languages and for multiple architecture targets.

With the increasing interest in IR to improve the performance of code generation models, new datasets are emerging. For example, ComPile~\citep{grossman2024compile}, a 2.4~TB dataset of unoptimized LLVM-IR.

\paragraph{Machine Learning in Compilers.} Many works have applied machine learning in compilers~\citep{ml4sysreview,Ashouri_2022,cummins_end2end,phothilimthana2021flexible,seeker2024revealing}. Compiler pass ordering has been exploited for decades. Over the years there have been several approaches using machine learning~\citep{coreset, agakov_iterative_2006,ogilvie_minimizing_2017,jayatilaka2021towards,queiroz2023graph,compiler-feedback}. Neural machine translation is an emerging field that uses language models to transform code from one language to another. Prior examples include compiling C to assembly~\citep{armengol2021learning}, assembly to C~\citep{armengol2023slade,hosseini2022beyond}, and source-to-source~\citep{transcoder}.
\section{Discussion}

In this paper, we introduced \foundationmodel, a novel family of large language models specifically designed to address the challenges of code and compiler optimization. By extending the capabilities of the foundational \codellama model, \foundationmodel provides a robust, pre-trained platform that significantly enhances the understanding and manipulation of compiler intermediate representations and assembly language.

We release \foundationmodel under a bespoke commercial license to facilitate widespread access and collaboration, enabling both academic researchers and industry practitioners to explore, modify, and extend the model according to their specific needs.

\subsection{Limitations}

We have shown that \foundationmodel performs well at compiler optimization tasks and has improved understanding of compiler representations and assembly code over prior works, but there are limitations. The main limitation is the finite sequence length of inputs (context window). \foundationmodel supports a 16k token context windows, but program codes may be far longer. For example, 67\% of MiBench translation units exceeded this context window when formatted as flag tuning prompts, shown in Table~\ref{tab:mibench-overview}. To mitigate this we split larger translation units into individual functions, though this limits the scope of optimization that can be performed, and still 18\% of the split translation units remain too large for the model to accept as input. Researchers are adopting ever-increasing context windows~\citep{ding2023longnet}, but finite context windows remain a common concern with LLMs.

A second limitation, common to all LLMs, is the accuracy of model outputs. Users of \foundationmodel are advised to assess their models using evaluation benchmarks specific to compilers. Given that compilers are not bug-free, any suggested compiler optimizations must be rigorously tested. When a model decompiles assembly code, its accuracy should be confirmed through round trip, manual inspection, or unit testing. For some applications LLM generations can be constrained to regular expressions~\citep{prioritysampling}, or combined with automatic verification to ensure correctness~\citep{llm-vectorizer}.

\clearpage
\bibliography{references}
\bibliographystyle{tmlr}

\clearpage
\appendix
\section{Acknowledgements}

We would like to express our gratitude to all the people who helped us carry out this project:

\begin{itemize}
    \item Our product and program management team:  Abhishek Charnalia, Claire Roberts
    \item Participants in the red teaming exercises: Shengye Wan, Aaron Grattafiori, Chris Rohlf, Joshua Saxe, Ivan Evtimov
    \item Our sanctions team: Olivia Scheiber, Samuel Selvan, Kyle Watson, Jenlain Scott, Margaret Warner
    \item Our marketing and comms team: Ana Paula Kirschner Mofarrej, Lisa Brown Jaloza, Nisha Deo, Uzma Saeed
    \item Our leadership team: Joelle Pinneau, Naila Murray, Kim Hazelwood, Joe Spisak, Ragavan Srinivasan
    \item All the members of the \codellama team: Sida Wang, Jade Copet, Yossef Adi, Olivier Duchenne, Quentin Carbonneaux, Jonas Gehring, Nicolas Usunier, Fabian Gloeckle, Taco Cohen, Kunhao Zheng, Sten Sootla, Vegard Mella, Pierre Chambon, Wes BOUAZIZ.
\end{itemize}
\vspace*{\fill}
\section{Prompts}

\subsection{Compiler emulation prompts}

\begin{figure}[H]
\setcaptiontype{lstlisting}
\small
\textbf{Prompt:}
\begin{lstlisting}
[INST] Give the LLVM-IR for the following code when optimized using opt -p 'module(default<Oz>)':

<code>; ModuleID = '<stdin>'
source_filename = "-"
target datalayout = "e-m:e-p270:32:32-p271:32:32-p272:64:64-i64:64-f80:128-n8:16:32:64-S128"
target triple = "x86_64-unknown-linux-gnu"

; Function Attrs: minsize nounwind optsize uwtable
define dso_local i32 @add_two(i32 noundef %0, i32 noundef %1) #0 {
  %3 = alloca i32, align 4
  %4 = alloca i32, align 4
  store i32 %0, ptr %3, align 4, !tbaa !5
  store i32 %1, ptr %4, align 4, !tbaa !5
  %5 = load i32, ptr %3, align 4, !tbaa !5
  %6 = load i32, ptr %4, align 4, !tbaa !5
  %7 = add nsw i32 %5, %6
  ret i32 %7
}

attributes #0 = { minsize nounwind optsize uwtable "min-legal-vector-width"="0" "no-trapping-math"="true" "stack-protector-buffer-size"="8" "target-cpu"="x86-64" "target-features"="+cmov,+cx8,+fxsr,+mmx,+sse,+sse2,+x87" "tune-cpu"="generic" }

!llvm.module.flags = !{!0, !1, !2, !3}
!llvm.ident = !{!4}

!0 = !{i32 1, !"wchar_size", i32 4}
!1 = !{i32 8, !"PIC Level", i32 2}
!2 = !{i32 7, !"PIE Level", i32 2}
!3 = !{i32 7, !"uwtable", i32 2}
!4 = !{!"clang version 17.0.6 (git@github.com:fairinternal/CodeGen.git b05db9bbf7a92019267416c1bb9996fe6134e3f1)"}
!5 = !{!6, !6, i64 0}
!6 = !{!"int", !7, i64 0}
!7 = !{!"omnipotent char", !8, i64 0}
!8 = !{!"Simple C/C++ TBAA"}
</code>

The input code has instruction count 8 and binary size 65 bytes. [/INST] 
\end{lstlisting}
\textbf{Answer:}
\begin{lstlisting}
The LLVM-IR will have instruction count 2 and binary sise 53 bytes:

<code>; ModuleID = '<stdin>'
source_filename = "-"
target datalayout = "e-m:e-p270:32:32-p271:32:32-p272:64:64-i64:64-f80:128-n8:16:32:64-S128"
target triple = "x86_64-unknown-linux-gnu"

; Function Attrs: minsize mustprogress nofree norecurse nosync nounwind optsize willreturn memory(none) uwtable
define dso_local i32 @add_two(i32 noundef %0, i32 noundef %1) local_unnamed_addr #0 {
  %3 = add nsw i32 %1, %0
  ret i32 %3
}

attributes #0 = { minsize mustprogress nofree norecurse nosync nounwind optsize willreturn memory(none) uwtable "min-legal-vector-width"="0" "no-trapping-math"="true" "stack-protector-buffer-size"="8" "target-cpu"="x86-64" "target-features"="+cmov,+cx8,+fxsr,+mmx,+sse,+sse2,+x87" "tune-cpu"="generic" }

!llvm.module.flags = !{!0, !1, !2, !3}
!llvm.ident = !{!4}

!0 = !{i32 1, !"wchar_size", i32 4}
!1 = !{i32 8, !"PIC Level", i32 2}
!2 = !{i32 7, !"PIE Level", i32 2}
!3 = !{i32 7, !"uwtable", i32 2}
!4 = !{!"clang version 17.0.6 (git@github.com:fairinternal/CodeGen.git b05db9bbf7a92019267416c1bb9996fe6134e3f1)"}
</code>
\end{lstlisting}
\caption{Prompt used for \emph{IR} $\rightarrow$ \emph{IR} compiler emulation.
\label{fig:compiler-emulation-prompt-ir}}
\end{figure}

\begin{figure}[H]
\setcaptiontype{lstlisting}
\small
\textbf{Prompt:}
\begin{lstlisting}
[INST] Give the assembly for the following code when optimized using opt -p 'module(default<Oz>)':

<code>; ModuleID = '<stdin>'
source_filename = "-"
target datalayout = "e-m:e-p270:32:32-p271:32:32-p272:64:64-i64:64-f80:128-n8:16:32:64-S128"
target triple = "x86_64-unknown-linux-gnu"

; Function Attrs: minsize nounwind optsize uwtable
define dso_local i32 @add_two(i32 noundef %0, i32 noundef %1) #0 {
  %3 = alloca i32, align 4
  %4 = alloca i32, align 4
  store i32 %0, ptr %3, align 4, !tbaa !5
  store i32 %1, ptr %4, align 4, !tbaa !5
  %5 = load i32, ptr %3, align 4, !tbaa !5
  %6 = load i32, ptr %4, align 4, !tbaa !5
  %7 = add nsw i32 %5, %6
  ret i32 %7
}

attributes #0 = { minsize nounwind optsize uwtable "min-legal-vector-width"="0" "no-trapping-math"="true" "stack-protector-buffer-size"="8" "target-cpu"="x86-64" "target-features"="+cmov,+cx8,+fxsr,+mmx,+sse,+sse2,+x87" "tune-cpu"="generic" }

!llvm.module.flags = !{!0, !1, !2, !3}
!llvm.ident = !{!4}

!0 = !{i32 1, !"wchar_size", i32 4}
!1 = !{i32 8, !"PIC Level", i32 2}
!2 = !{i32 7, !"PIE Level", i32 2}
!3 = !{i32 7, !"uwtable", i32 2}
!4 = !{!"clang version 17.0.6 (git@github.com:fairinternal/CodeGen.git b05db9bbf7a92019267416c1bb9996fe6134e3f1)"}
!5 = !{!6, !6, i64 0}
!6 = !{!"int", !7, i64 0}
!7 = !{!"omnipotent char", !8, i64 0}
!8 = !{!"Simple C/C++ TBAA"}
</code>

The input code has instruction count 8 and binary size 65 bytes. [/INST] 
\end{lstlisting}
\textbf{Answer:}
\begin{lstlisting}
The assembly will have instruction count 2 and binary sise 53 bytes:

<code>	.text
	.file	"-"
	.globl	add_two                         # -- Begin function add_two
	.type	add_two,@function
add_two:                                # @add_two
	.cfi_startproc
# %bb.0:
	movl	%esi, %eax
	addl	%edi, %eax
	retq
.Lfunc_end0:
	.size	add_two, .Lfunc_end0-add_two
	.cfi_endproc
                                        # -- End function
	.ident	"clang version 17.0.6 (git@github.com:fairinternal/CodeGen.git b05db9bbf7a92019267416c1bb9996fe6134e3f1)"
	.section	".note.GNU-stack","",@progbits
	.addrsig
</code>
\end{lstlisting}
\caption{Prompt used for \emph{IR} $\rightarrow$ \emph{assembly} compiler emulation.
\label{fig:compiler-emulation-prompt-asm}}
\end{figure}

\clearpage
\subsection{Flag tuning prompts}

\begin{figure}[H]
\setcaptiontype{lstlisting}
\small
\textbf{Prompt:}
\begin{lstlisting}
[INST] Tell me how to optimize this LLVM-IR for object file size:

<code>; ModuleID = '<stdin>'
source_filename = "-"
target datalayout = "e-m:e-p270:32:32-p271:32:32-p272:64:64-i64:64-f80:128-n8:16:32:64-S128"
target triple = "x86_64-unknown-linux-gnu"

; Function Attrs: minsize nounwind optsize uwtable
define dso_local i32 @add_two(i32 noundef %0, i32 noundef %1) #0 {
  %3 = alloca i32, align 4
  %4 = alloca i32, align 4
  store i32 %0, ptr %3, align 4, !tbaa !5
  store i32 %1, ptr %4, align 4, !tbaa !5
  %5 = load i32, ptr %3, align 4, !tbaa !5
  %6 = load i32, ptr %4, align 4, !tbaa !5
  %7 = add nsw i32 %5, %6
  ret i32 %7
}

attributes #0 = { minsize nounwind optsize uwtable "min-legal-vector-width"="0" "no-trapping-math"="true" "stack-protector-buffer-size"="8" "target-cpu"="x86-64" "target-features"="+cmov,+cx8,+fxsr,+mmx,+sse,+sse2,+x87" "tune-cpu"="generic" }

!llvm.module.flags = !{!0, !1, !2, !3}
!llvm.ident = !{!4}

!0 = !{i32 1, !"wchar_size", i32 4}
!1 = !{i32 8, !"PIC Level", i32 2}
!2 = !{i32 7, !"PIE Level", i32 2}
!3 = !{i32 7, !"uwtable", i32 2}
!4 = !{!"clang version 17.0.6 (git@github.com:fairinternal/CodeGen.git b05db9bbf7a92019267416c1bb9996fe6134e3f1)"}
!5 = !{!6, !6, i64 0}
!6 = !{!"int", !7, i64 0}
!7 = !{!"omnipotent char", !8, i64 0}
!8 = !{!"Simple C/C++ TBAA"}</code> [/INST]
\end{lstlisting}
\textbf{Answer:}
\begin{lstlisting}
The code has object file size of 65 which can be reduced to 53 by running:

`opt -p 'module(default<Oz>)'`

This will produce code:

<code>; ModuleID = '<stdin>'
source_filename = "-"
target datalayout = "e-m:e-p270:32:32-p271:32:32-p272:64:64-i64:64-f80:128-n8:16:32:64-S128"
target triple = "x86_64-unknown-linux-gnu"

; Function Attrs: minsize mustprogress nofree norecurse nosync nounwind optsize willreturn memory(none) uwtable
define dso_local i32 @add_two(i32 noundef %0, i32 noundef %1) local_unnamed_addr #0 {
  %3 = add nsw i32 %1, %0
  ret i32 %3
}

attributes #0 = { minsize mustprogress nofree norecurse nosync nounwind optsize willreturn memory(none) uwtable "min-legal-vector-width"="0" "no-trapping-math"="true" "stack-protector-buffer-size"="8" "target-cpu"="x86-64" "target-features"="+cmov,+cx8,+fxsr,+mmx,+sse,+sse2,+x87" "tune-cpu"="generic" }

!llvm.module.flags = !{!0, !1, !2, !3}
!llvm.ident = !{!4}

!0 = !{i32 1, !"wchar_size", i32 4}
!1 = !{i32 8, !"PIC Level", i32 2}
!2 = !{i32 7, !"PIE Level", i32 2}
!3 = !{i32 7, !"uwtable", i32 2}
!4 = !{!"clang version 17.0.6 (git@github.com:fairinternal/CodeGen.git b05db9bbf7a92019267416c1bb9996fe6134e3f1)"}</code>
\end{lstlisting}
\caption{Prompt used for flag tuning.
\label{fig:flag-tuning-prompt}}
\end{figure}

\begin{figure}[H]
\setcaptiontype{lstlisting}
\small
\textbf{Prompt:}
\begin{lstlisting}
[INST] Tell me how to optimize this LLVM-IR for object file size:

<code>; ModuleID = '<stdin>'
source_filename = "-"
target datalayout = "e-m:e-p270:32:32-p271:32:32-p272:64:64-i64:64-f80:128-n8:16:32:64-S128"
target triple = "x86_64-unknown-linux-gnu"

; Function Attrs: minsize mustprogress nofree norecurse nosync nounwind optsize willreturn memory(none) uwtable
define dso_local i32 @add_two(i32 noundef %0, i32 noundef %1) local_unnamed_addr #0 {
  %3 = add nsw i32 %1, %0
  ret i32 %3
}

attributes #0 = { minsize mustprogress nofree norecurse nosync nounwind optsize willreturn memory(none) uwtable "min-legal-vector-width"="0" "no-trapping-math"="true" "stack-protector-buffer-size"="8" "target-cpu"="x86-64" "target-features"="+cmov,+cx8,+fxsr,+mmx,+sse,+sse2,+x87" "tune-cpu"="generic" }

!llvm.module.flags = !{!0, !1, !2, !3}
!llvm.ident = !{!4}

!0 = !{i32 1, !"wchar_size", i32 4}
!1 = !{i32 8, !"PIC Level", i32 2}
!2 = !{i32 7, !"PIE Level", i32 2}
!3 = !{i32 7, !"uwtable", i32 2}
!4 = !{!"clang version 17.0.6 (git@github.com:fairinternal/CodeGen.git b05db9bbf7a92019267416c1bb9996fe6134e3f1)"}</code> [/INST] 
\end{lstlisting}
\textbf{Answer:}
\begin{lstlisting}
The code has object file size of 53 which cannot be reduced further
\end{lstlisting}
\caption{Prompt and answer format for flag tuning task when no improvement can be made.
\label{fig:flag-tuning-prompt-no-improvement}}
\end{figure}

\begin{figure}[H]
\setcaptiontype{lstlisting}
\small
\begin{lstlisting}
Provide your answer as a list of command line options to opt version 17.0.6, using the format:

"$ opt -p '<passes>'"

Where <passes> is a list of passes for the new pass manager, e.g. "function(dce),module(default<Oz>),function(load-store-vectorizer)".

Only include the passes list. Do not include file paths or other flags such as -o. Terminate the opt command line options with a newline.

Then report the optimized code that will be produced, delimited by <code> and </code> tags.

Finally, report the binary size the code before and after optimization using the template:

"Before optimization: X bytes. After optimization: Y bytes."

Where X and Y are placeholders for integer binary sizes in bytes. Binary size is the summation of the .text and .data segment sizes of the object file generated by `clang-17 output.bc -c`, as reported by the `size` tool.

Include no other text in your response.
\end{lstlisting}
\caption{Additional suffix appended to flag tuning prompts when evaluating GPT-4 and \codellamainst. We found that this additional context helped improve the coherence of model responses.
\label{fig:flag-tuning-prompt-suffix}}
\end{figure}

\subsection{Disassembly prompts}

\begin{figure}[H]
\setcaptiontype{lstlisting}
\small
\textbf{Prompt:}
\begin{lstlisting}
[INST] Disassemble this code to LLVM-IR:

<code>	.text
	.file	"-"
	.globl	add_two                         # -- Begin function add_two
	.type	add_two,@function
add_two:                                # @add_two
	.cfi_startproc
# %bb.0:
	movl	%edi, -4(%rsp)
	movl	%esi, -8(%rsp)
	movl	-4(%rsp), %eax
	addl	-8(%rsp), %eax
	retq
.Lfunc_end0:
	.size	add_two, .Lfunc_end0-add_two
	.cfi_endproc
                                        # -- End function
	.ident	"clang version 17.0.6 (git@github.com:fairinternal/CodeGen.git b05db9bbf7a92019267416c1bb9996fe6134e3f1)"
	.section	".note.GNU-stack","",@progbits
	.addrsig
</code> [/INST] 
\end{lstlisting}
\textbf{Answer:}
\begin{lstlisting}
<code>; ModuleID = '<stdin>'
source_filename = "-"
target datalayout = "e-m:e-p270:32:32-p271:32:32-p272:64:64-i64:64-f80:128-n8:16:32:64-S128"
target triple = "x86_64-unknown-linux-gnu"

; Function Attrs: minsize nounwind optsize uwtable
define dso_local i32 @add_two(i32 noundef %0, i32 noundef %1) #0 {
  %3 = alloca i32, align 4
  %4 = alloca i32, align 4
  store i32 %0, ptr %3, align 4, !tbaa !5
  store i32 %1, ptr %4, align 4, !tbaa !5
  %5 = load i32, ptr %3, align 4, !tbaa !5
  %6 = load i32, ptr %4, align 4, !tbaa !5
  %7 = add nsw i32 %5, %6
  ret i32 %7
}

attributes #0 = { minsize nounwind optsize uwtable "min-legal-vector-width"="0" "no-trapping-math"="true" "stack-protector-buffer-size"="8" "target-cpu"="x86-64" "target-features"="+cmov,+cx8,+fxsr,+mmx,+sse,+sse2,+x87" "tune-cpu"="generic" }

!llvm.module.flags = !{!0, !1, !2, !3}
!llvm.ident = !{!4}

!0 = !{i32 1, !"wchar_size", i32 4}
!1 = !{i32 8, !"PIC Level", i32 2}
!2 = !{i32 7, !"PIE Level", i32 2}
!3 = !{i32 7, !"uwtable", i32 2}
!4 = !{!"clang version 17.0.6 (git@github.com:fairinternal/CodeGen.git b05db9bbf7a92019267416c1bb9996fe6134e3f1)"}
!5 = !{!6, !6, i64 0}
!6 = !{!"int", !7, i64 0}
!7 = !{!"omnipotent char", !8, i64 0}
!8 = !{!"Simple C/C++ TBAA"}
</code>
\end{lstlisting}
\caption{Prompt used for disassembly.
\label{fig:disassembly-prompt}}
\end{figure}

\begin{figure}[H]
\setcaptiontype{lstlisting}
\small
\begin{lstlisting}
Use LLVM version 17.0.6. Provide the IR enclosed by <code> and </code> tags.

Include no other text.
\end{lstlisting}
\caption{Additional suffix appended to disassembly prompts when evaluating GPT-4 and \codellamainst. We found that this additional context helped improve the coherence of model responses.
\label{fig:disassembly-prompt-suffix}}
\end{figure}

\section{Optimization Passes}

\begin{table}[H]
\caption{List of \textit{opt} 17.0.6 transformation passes used to generate data for code optimization self-training and flag-tuning tasks.}
{\scriptsize
\begin{tabular}{lc|lc|lc}
\toprule
Pass Name & Level & Pass Name & Level & Pass Name & Level \\
\midrule
O0 & Module & no-op-cgscc & CGSCC & lcssa & Function \\
O1 & Module & inline & CGSCC & loop-data-prefetch & Function \\
O2 & Module & coro-split & CGSCC & loop-load-elim & Function \\
O3 & Module & function-attrs & CGSCC & loop-fusion & Function \\
Os & Module & aa-eval & Function & loop-distribute & Function \\
Oz & Module & adce & Function & loop-versioning & Function \\
always-inline & Module & add-discriminators & Function & pa-eval & Function \\
attributor & Module & aggressive-instcombine & Function & place-safepoints & Function \\
annotation2metadata & Module & assume-builder & Function & reassociate & Function \\
openmp-opt & Module & assume-simplify & Function & redundant-dbg-inst-elim & Function \\
openmp-opt-postlink & Module & alignment-from-assumptions & Function & reg2mem & Function \\
called-value-propagation & Module & annotation-remarks & Function & scalarize-masked-mem-intrin & Function \\
canonicalize-aliases & Module & bdce & Function & scalarizer & Function \\
constmerge & Module & break-crit-edges & Function & separate-const-offset-from-gep & Function \\
coro-early & Module & callsite-splitting & Function & sccp & Function \\
coro-cleanup & Module & consthoist & Function & sink & Function \\
cross-dso-cfi & Module & count-visits & Function & slp-vectorizer & Function \\
deadargelim & Module & constraint-elimination & Function & slsr & Function \\
elim-avail-extern & Module & chr & Function & speculative-execution & Function \\
extract-blocks & Module & coro-elide & Function & strip-gc-relocates & Function \\
forceattrs & Module & correlated-propagation & Function & tailcallelim & Function \\
globalopt & Module & dce & Function & vector-combine & Function \\
globalsplit & Module & dfa-jump-threading & Function & tlshoist & Function \\
hotcoldsplit & Module & div-rem-pairs & Function & declare-to-assign & Function \\
inferattrs & Module & dse & Function & early-cse & Function \\
inliner-wrapper & Module & fix-irreducible & Function & ee-instrument & Function \\
inliner-wrapper-no-mandatory-first & Module & flattencfg & Function & hardware-loops & Function \\
iroutliner & Module & make-guards-explicit & Function & lower-matrix-intrinsics & Function \\
lower-global-dtors & Module & gvn-hoist & Function & loop-unroll & Function \\
lower-ifunc & Module & gvn-sink & Function & simplifycfg & Function \\
lowertypetests & Module & infer-address-spaces & Function & loop-vectorize & Function \\
mergefunc & Module & instcombine & Function & instcombine & Function \\
name-anon-globals & Module & instsimplify & Function & mldst-motion & Function \\
partial-inliner & Module & irce & Function & gvn & Function \\
recompute-globalsaa & Module & float2int & Function & sroa & Function \\
rel-lookup-table-converter & Module & libcalls-shrinkwrap & Function & loop-flatten & Loop \\
rewrite-statepoints-for-gc & Module & inject-tli-mappings & Function & loop-interchange & Loop \\
rewrite-symbols & Module & instnamer & Function & loop-unroll-and-jam & Loop \\
rpo-function-attrs & Module & lower-expect & Function & canon-freeze & Loop \\
scc-oz-module-inliner & Module & lower-guard-intrinsic & Function & loop-idiom & Loop \\
strip & Module & lower-constant-intrinsics & Function & loop-instsimplify & Loop \\
strip-dead-debug-info & Module & lower-widenable-condition & Function & loop-deletion & Loop \\
strip-dead-prototypes & Module & guard-widening & Function & loop-simplifycfg & Loop \\
strip-debug-declare & Module & load-store-vectorizer & Function & loop-reduce & Loop \\
strip-nondebug & Module & loop-simplify & Function & indvars & Loop \\
strip-nonlinetable-debuginfo & Module & loop-sink & Function & loop-unroll-full & Loop \\
synthetic-counts-propagation & Module & lowerswitch & Function & loop-predication & Loop \\
wholeprogramdevirt & Module & mem2reg & Function & guard-widening & Loop \\
module-inline & Module & memcpyopt & Function & loop-bound-split & Loop \\
pseudo-probe-update & Module & mergeicmps & Function & loop-reroll & Loop \\
globaldce & Module & mergereturn & Function & loop-versioning-licm & Loop \\
ipsccp & Module & move-auto-init & Function & simple-loop-unswitch & Loop \\
embed-bitcode & Module & nary-reassociate & Function & loop-rotate & Loop \\
argpromotion & CGSCC & newgvn & Function & licm & LoopMssa \\
attributor-cgscc & CGSCC & jump-threading & Function & lnicm & LoopMssa \\
openmp-opt-cgscc & CGSCC & partially-inline-libcalls & Function & & \\
\bottomrule
\end{tabular}
}
\label{tab:passes}
\end{table}

\begin{figure}[H]
\begin{center}
\centerline{\includegraphics[width=1\textwidth]{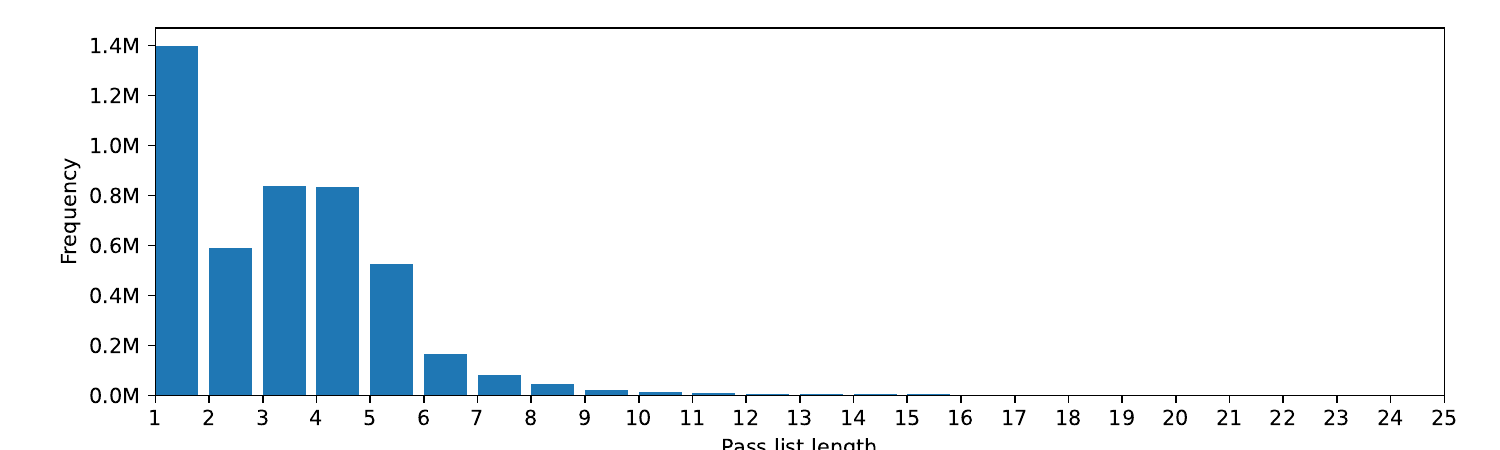}}
\caption{%
    Length of autotuned pass lists.
}
\label{fig:pass-list-len}
\end{center}
\end{figure}

\begin{figure}[H]
\begin{center}
\centerline{\includegraphics[width=.55\textwidth]{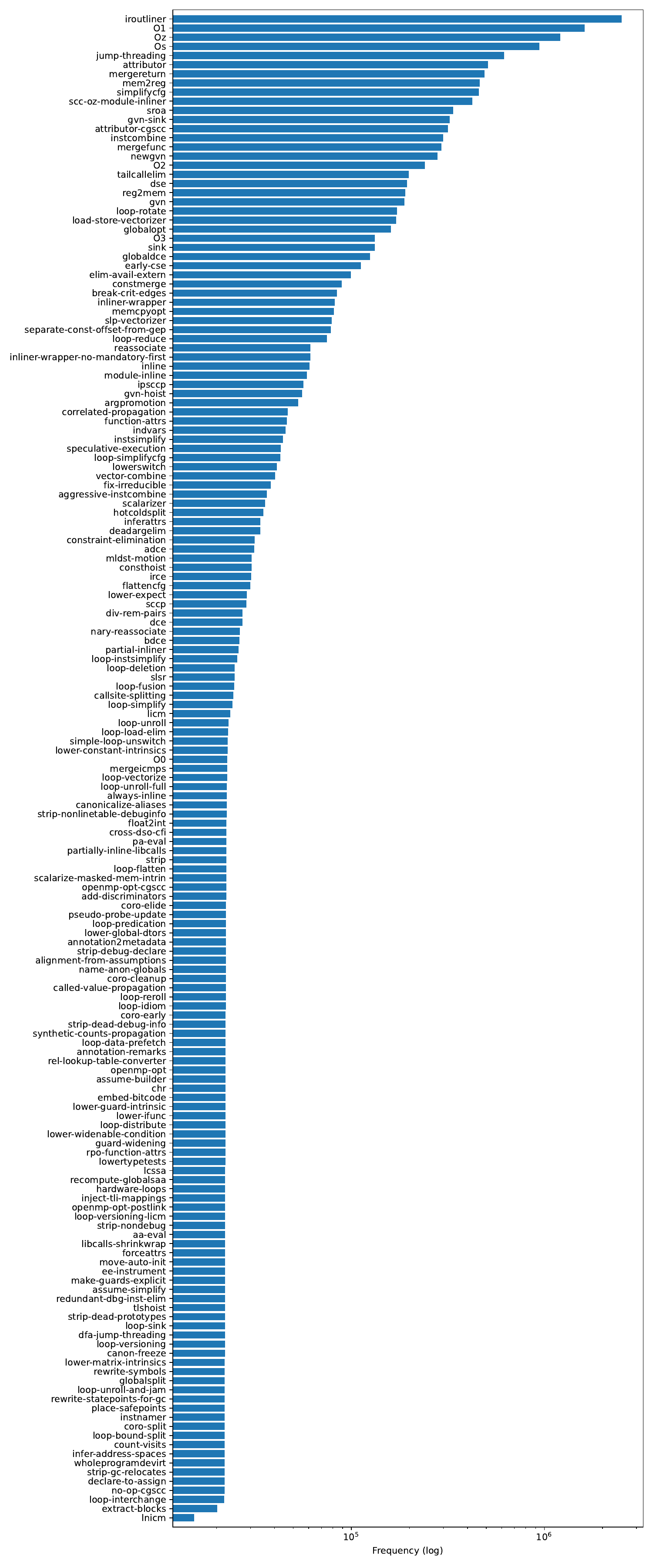}}
\caption{%
    The frequency of passes in the best pass lists generated by the autotuner on our training programs.
    \label{fig:autotuner-best-pass-freq}
}
\end{center}
\end{figure}

\section{Benchmarks}

\begin{table}[H]
\small
\caption{%
  MiBench benchmarks used for flag tuning task evaluation.%
  \label{tab:mibench-overview}%
}
\begin{tabular}{lccC{3cm}C{2.5cm}C{3cm}}
\toprule
 & & \multicolumn{2}{c}{Without split} & \multicolumn{2}{c}{With split} \\
 & Binary size &  Translation units & Truncated prompts & Translation units & Truncated prompts \\
\midrule
adpcm        &    816.7 kB &         2 &              &                 2 &                        \\
basicmath    &    931.7 kB &         4 &              &                 4 &                        \\
bitcount     &    821.1 kB &         8 &              &                 8 &                        \\
blowfish     &    830.6 kB &         7 &      3 (43\%) &                 7 &                2 (29\%) \\
crc32        &    818.4 kB &         1 &              &                 1 &                        \\
dijkstra     &    946.0 kB &         1 &              &                 1 &                        \\
fft          &    844.8 kB &         3 &              &                 3 &                        \\
ghostscript  &      1.9 MB &       296 &    222 (75\%) &              1,052 &              162 (15\%) \\
gsm          &     58.8 kB &        23 &     12 (52\%) &                37 &               10 (27\%) \\
ispell       &     91.5 kB &        12 &      8 (67\%) &                39 &                6 (15\%) \\
jpeg\_c       &    112.5 kB &        54 &     39 (72\%) &               170 &               22 (13\%) \\
jpeg\_d       &    151.7 kB &        54 &     39 (72\%) &               164 &               18 (11\%) \\
lame         &    289.2 kB &        32 &     22 (69\%) &                92 &               24 (26\%) \\
patricia     &    949.3 kB &         2 &      1 (50\%) &                 3 &                        \\
qsort        &    944.3 kB &         1 &              &                 1 &                        \\
rsynth       &    151.4 kB &        19 &     10 (53\%) &                27 &                3 (11\%) \\
sha          &      5.3 kB &         2 &      1 (50\%) &                 3 &                        \\
stringsearch &    821.5 kB &         4 &              &                 4 &                        \\
susan        &    911.4 kB &         1 &     1 (100\%) &                13 &                7 (54\%) \\
tiff2bw      &    442.1 kB &        34 &     19 (56\%) &               134 &               24 (18\%) \\
tiff2rgba    &    492.7 kB &        34 &     19 (56\%) &               134 &               23 (17\%) \\
tiffdither   &    441.2 kB &        34 &     19 (56\%) &               133 &               23 (17\%) \\
tiffmedian   &    453.0 kB &        34 &     19 (56\%) &               139 &               26 (19\%) \\
typeset      &      2.0 MB &        51 &     43 (84\%) &               227 &               89 (39\%) \\
\midrule
Total & & 713 & 477 (67\%) & 2,398 & 439 (18\%) \\
\bottomrule
\end{tabular}
\end{table}

\begin{table}[H]
\small
\centering
\caption{MiBench benchmarks used for disassembly task evaluation. \label{tab:mibench-disassembly-overview}}
\begin{tabular}{lcc}
\toprule
{} & Translation units & Truncated prompts \\
\midrule
adpcm        &                 3 &                   \\
basicmath    &                 2 &                   \\
bitcount     &                 8 &                   \\
blowfish     &                 3 &                   \\
crc32        &                 1 &                   \\
dijkstra     &                 2 &                   \\
fft          &                 1 &                   \\
ghostscript  &             1,264 &                 2 \\
gsm          &                35 &                   \\
ispell       &                45 &                   \\
jpeg\_c       &                24 &                   \\
jpeg\_d       &               177 &                   \\
lame         &                87 &                 1 \\
patricia     &                 3 &                   \\
qsort        &                 1 &                   \\
rsynth       &                33 &                 1 \\
sha          &                 3 &                   \\
stringsearch &                 5 &                   \\
susan        &                 7 &                   \\
tiff2bw      &                 3 &                   \\
tiff2rgba    &                 5 &                   \\
tiffdither   &                 2 &                   \\
tiffmedian   &               158 &                   \\
typeset      &               143 &                   \\
\midrule
Total & 2015 & 4 \\
\bottomrule
\end{tabular}
\end{table}

\newpage\clearpage

\section{Model card}

\begin{table}[h!]
{
\small
    \centering
    \begin{tabular}{p{0.18\textwidth}|l}
    \thickhline
    \multicolumn{2}{c}{\bf Model details}\\
    \thickhline
        Model Developers & Meta AI 
        \\\hline
        Variations & \makecell[lp{0.75\textwidth}]{
        \foundationmodel comes in two model sizes: 7B and 13B parameters. Both variations have been trained on the same data. \model, available in the same sizes, extends these with further training.
        }\\\hline
        Input&Models input text only.
        \\\hline
        Output&Models output text only.
        \\\hline
        Model Architecture&\makecell[lp{0.75\textwidth}]{
        \foundationmodel and its variants are autoregressive language models using optimized transformer architectures. All models were fine-tuned with up to 16K tokens.
        }\\\hline
        Model Dates&\makecell[lp{0.75\textwidth}]{
        \foundationmodel and its variants have been trained between January and May 2024.
        } 
        \\\hline 
        Status&\makecell[lp{0.75\textwidth}]{
        This is a static model trained on an offline dataset.}
        \\\hline 
        Licence&\makecell[lp{0.75\textwidth}]{
        A custom commercial license is available at: \url{ai.meta.com/resources/models-and-libraries/llama-downloads/}.
        }\\\hline 
        \makecell[lp{0.18\textwidth}]{Where to send comments}&\makecell[lp{0.75\textwidth}]{
        Instructions on how to provide feedback or comments on the model can be found in the model README.}
        \\
        \thickhline
        \multicolumn{2}{c}{\bf Intended Use}\\
        \thickhline
        Intended Use Cases&\makecell[lp{0.75\textwidth}]{
        \foundationmodel and its variants are intended for commercial and research use in English and relevant programming languages. 
        The foundation model \foundationmodel can be adapted for a variety of code optimization and understanding tasks. 
        }\\\hline 
        Out-of-Scope Uses&\makecell[lp{0.75\textwidth}]{
        Use in any manner that violates applicable laws or regulations (including trade compliance laws). Use in languages other than English. Use in any other way that is prohibited by the Acceptable Use Policy and Licensing Agreement for \foundationmodel and its variants.
        }\\
        \thickhline
        \multicolumn{2}{c}{\bf Hardware and Software}\\
        \thickhline
        Training Factors&\makecell[lp{0.75\textwidth}]{
        We used custom training libraries. The training and fine-tuning of the released models have been performed on Meta’s Research Super Cluster.
        }\\\hline 
        Carbon Footprint&\makecell[lp{0.75\textwidth}]{
        In aggregate, training all 4 \foundationmodel models required 264K GPU hours of computation on hardware of type A100-80GB (TDP of 350-400W). 
        Estimated total emissions were 64.12 tCO2eq, 100\% of which were offset by Meta’s sustainability program.
        }\\
        \thickhline
        \multicolumn{2}{c}{\bf Training Data}\\
        \thickhline
        \multicolumn{2}{c}{
        \makecell[lp{0.95\textwidth}]{
        All experiments reported here and the released models have been trained and fine-tuned using the same data as \codellama with different weights (see Section~\ref{sec:method} and Table~\ref{tab:dataset}).
        }
        }\\\thickhline 
        \multicolumn{2}{c}{\bf Evaluation Results}\\
        \thickhline
        \multicolumn{2}{c}{
        \makecell[lp{0.95\textwidth}]{See evaluations for the main models and detailed ablations \Cref{sec:results}.
        }}
        \\\thickhline
        \multicolumn{2}{c}{\bf Ethical Considerations and Limitations}
        \\\thickhline
        \multicolumn{2}{c}{
        \makecell[lp{0.95\textwidth}]{\foundationmodel and its variants are a new technology that carries risks with use. Testing conducted to date has been in English, and has not covered, nor could it cover all scenarios. For these reasons, as with all LLMs, \foundationmodel’s potential outputs cannot be predicted in advance, and the model may in some instances produce inaccurate or objectionable responses to user prompts. Therefore, before deploying any applications of \foundationmodel, developers should perform safety testing and tuning tailored to their specific applications of the model. Please see the Responsible Use Guide available available at \url{https://ai.meta.com/llama/responsible-user-guide}.}}
        \\
        \thickhline
    \end{tabular}
    \caption{Model card~\citep{mitchell2019modelcard} for \foundationmodel and \model.}
    \label{tab:modelcard}
}
\end{table}

\end{document}